\documentclass[12pt,a4paper,fleqn]{article}
\usepackage[textheight=23cm,textwidth=15cm]{geometry}
\usepackage{amsmath}
\usepackage{graphicx}
\usepackage{cite}
\usepackage[american]{babel}
\usepackage{here}
\usepackage{fixltx2e} 

\begin{document}

\begin{center}

{\LARGE\bf
 Statistical Analysis of Semiclassical Dispersion Corrections
}

\vspace{1cm}

{\large
Thomas Weymuth$^{a}$,
Jonny Proppe$^{a}$,
and Markus Reiher$^{a,}$\footnote{Corresponding author; e-mail: markus.reiher@phys.chem.ethz.ch}
}\\[4ex]

$^{a}$ Laboratorium f\"ur Physikalische Chemie, ETH Z\"urich, \\
Vladimir-Prelog-Weg 2, 8093 Z\"urich, Switzerland

January 25, 2018

\vspace{.41cm}

\end{center}

\begin{center}
\textbf{Abstract}
\end{center}
\vspace*{-.41cm}
{\small
Semiclassical dispersion corrections developed by Grimme and coworkers have become indispensable in applications of Kohn--Sham density functional theory.
A deeper understanding of the underlying parameterization might be crucial for well-founded further improvements of this successful approach.
To this end, we present an in-depth assessment of the fit parameters present in semiclassical (D3-type) dispersion corrections by means of a statistically rigorous analysis. 
We find that the choice of the cost function generally has a small effect on the empirical parameters of D3-type dispersion corrections with respect to the reference 
set under consideration.
Only in a few cases, the choice of cost function has a surprisingly large effect on the total dispersion energies.
In particular, the weighting scheme in the cost function can significantly affect the reliability of predictions.
In order to obtain unbiased (data-independent) uncertainty estimates for both the empirical fit parameters and the corresponding predictions, we carried out a nonparametric 
bootstrap analysis. 
This analysis reveals that the standard deviation of the mean of the empirical D3 parameters is small. Moreover, the mean prediction uncertainty obtained by bootstrapping 
is not much larger than previously reported error measures. 
On the basis of a jackknife analysis, we find that the original reference set is slightly skewed, but our results also suggest that this feature 
hardly affects the prediction of dispersion energies.
Furthermore, we find that the introduction of small uncertainties to the reference data does not change the conclusions drawn in this paper.
However, a rigorous analysis of error accumulation arising from different parameterizations reveals that error cancellation does not necessarily occur, leading to a monotonically 
increasing deviation in the dispersion energy with increasing molecule size.
We discuss this issue in detail at the prominent example of the C$_{60}$ ``buckycatcher''.
We find deviations between individual parameterizations of several tens of kcal\,mol$^{-1}$ in some cases.
Hence, in combination with any calculation of dispersion energies, we recommend to always determine the associated uncertainties
for which we will provide a software tool.
}

\section{Introduction}
\label{sec:intro}

Over the last decades, density functional theory (DFT) and in  particular Kohn--Sham DFT (KS-DFT)\cite{phys_rev_1965_140_a1133} 
has become the major workhorse for a broad range of applications in computational and theoretical chemistry, because it offers a satisfactory trade-off between the 
accuracy achievable and the computational resources required\cite{kohn1999, koch2001, theory_and_appl_of_comp_chem,
j_chem_phys_2014_140_18a301}. 
However, long-range electron correlation, and hence, London dispersion interaction, are not described by most exchange--correlation functionals 
(see, \textit{e.g.}, Refs.~\cite{j_phys_chem_a_2005_109_11015, chem_phys_lett_2006_419_333, chem_phys_lett_1994_229_175, chem_phys_lett_1995_233_134,wodrich2006}). 
This problem has been addressed in a number of ways\cite{j_phys_org_chem_2009_22_1127, j_chem_phys_2012_137_120901, j_phys_condens_matter_2012_24_073201,
 adv_funct_mater_2015_25_2054, rev_mod_phys_2010_82_1887, chem_rev_2016_116_5105}. 
For example, Corminboeuf and coworkers developed a density-dependent dispersion correction for KS-DFT applicable to the entire periodic table \cite{steinmann2010}.
Roethlisberger and coworkers designed atom-centered, nonlocal potentials to take into account noncovalent interactions\cite{phys_rev_lett_2004_93_153004}.
Truhlar and coworkers designed exchange--correlation functionals which are highly parameterized to describe weak noncovalent 
interactions better\cite{phil_trans_r_soc_a_2014_372_20120476}, whereas Dion \textit{et al.}~constructed a functional form which includes dispersion interactions in a 
nonempirical way\cite{phys_rev_lett_2004_92_246401, phys_rev_lett_2005_95_109902}.
In this respect, the nonlocal correlation energy functional by Van Voorhis and coworkers, where the number of empirical parameters is kept to a 
minimum\cite{vydrov2010}, can be considered an intermediate between the former approaches.

To date, the most widely used approach is the introduction of semiclassical dispersion corrections, which are added to the electronic energy \textit{a posteriori}.
They have been proposed as early as the 1970s\cite{j_chem_phys_1974_61_2372, ahlrichs1977}.
Significant advances of this direction were achieved in the 2000s by Becke and Johnson\cite{j_chem_phys_2005_123_154101, j_chem_phys_2005_123_024101, j_chem_phys_2006_124_174104}, 
by Tkatchenko and Scheffler\cite{phys_rev_lett_2009_102_073005, phys_rev_lett_2012_108_236402}, and by Grimme and coworkers\cite{j_comp_chem_2004_25_1463,
j_comp_chem_2006_27_1787,grimme2010,grimme2011,j_chem_phys_2017_147_034112} (D$x$-type dispersion corrections).
Such corrections are designed to avoid double-counting of electron correlation effects.
Especially the so-called D3 approach by Grimme \cite{grimme2010, grimme2011} has found widespread use as is evident from bibliometric tools such as Web of Science
and Google Scholar.
Therefore, D3 is the \textit{de facto} standard for addressing nonlocal interactions in KS-DFT.

Like almost all parameterized physicochemical property models (\textit{e.g.}, approximations to the exact exchange--correlation functional), 
semiclassical dispersion corrections have not been subjected to a rigorous statistical analysis yet.
Here, we present such an analysis of D3-type dispersion corrections with a focus on parameter optimization, reference set selection, 
uncertainty propagation, and model transferability.
All D$x$-type dispersion corrections ($x \in \lbrace 2,3,4 \rbrace$) comprise a small number (usually two or three) of global empirical parameters (apart from the $C_n^{AB}$ 
coefficients that account for pair-specific contributions; see below) that are calibrated against a 
set of reference dispersion energies.
In this work, we consider the D3-zero and D3-BJ dispersion corrections (\textit{i.e.}, D3 models with a zero-damping function\cite{grimme2010} and 
a Becke--Johnson (BJ) damping function\cite{grimme2011}, respectively), the latter being the most widely applied D3 variant.
It is advisable to subject the corresponding reference sets (called fit sets in the original literature\cite{grimme2010, grimme2011}) as well as the  
calibration procedure to a detailed statistical analysis to estimate the (co)variance of the empirical parameters and its effect on the prediction of dispersion energies.

Among other things, this strategy will allow us to estimate reliably the model prediction uncertainty (MPU), \textit{i.e.}, the expected (random) deviation 
of a prediction from a measurement or computational result\,---\,as of now, the MPU of semiclassical dispersion corrections has usually been assessed by calculating 
some error measure for a given validation set. 
Such MPU measures, however, have a strong dependence on the  validation set chosen.
The tools and methods required for an improved statistical analysis have only recently been embraced in quantum chemistry\cite{irikura2004, mortensen2005, 
sutton2016, simm2016, proppe2016, simm2017, proppe2017, pernot2015, pernot2017, pernot2017a, pernot2018, oung2018}. 
In this study, we revise the original calibration procedure of the D3 approach, and employ nonparametric bootstrapping\cite{ann_statist_1979_7_1} to obtain 
reliable estimates for both the uncertainty of the D3-BJ parameters and the corresponding MPU that is approximately independent 
of the validation set chosen.

We should note that the fourth generation of dispersion corrections by Grimme and coworkers (D4) has recently been published\cite{j_chem_phys_2017_147_034112}. 
We deliberately chose the D3-BJ model as the object of our analysis because of its importance in current computational studies. We may also note that revised 
parameterizations for D3-zero and D3-BJ have been developed\cite{j_phys_chem_lett_2016_7_2197}, which, however, have never attracted as much attention as the 
original parameterization.
Furthermore, since the dispersion corrections by Grimme and coworkers feature the same functional form for a given damping function (\textit{e.g.}, D3-BJ and D4-BJ only 
differ in the values of the dispersion coefficients $C_n^{AB}$, see below), one can expect our results to be generalizable.

This paper is organized as follows: 
In Section~\ref{sec:comp_meth}, we detail the methodology adopted in this study. 
In Section~\ref{sec:params}, we provide a detailed analysis of the D3-zero and D3-BJ parameterizations.
We further examine the effect of different weighting schemes on the D3-BJ parameterization in Section~\ref{sec:weighting}.
In Section~\ref{sec:bootstrap}, we demonstrate how to infer the uncertainty of the empirical D3-BJ parameters on the basis of a bootstrap analysis, before we harness 
the related jackknife method in Section~\ref{sec:jackknife} to identify reference data that bias the D3-BJ  parameterization.
Subsequently, in Section~\ref{sec:reference}, we discuss the effect of uncertainty in the reference data on the parameterization; an issue that is 
disregarded in most quantum chemical calibration studies.
Finally, the insights obtained will serve to provide an in-depth assessment of the molecule-size dependence of the model prediction uncertainty of the dispersion 
corrections in Section~\ref{sec:size}.

\section{Computational Methodology}
\label{sec:comp_meth}

The generic D3 model by Grimme \cite{grimme2010, grimme2011} is an additive \textit{a posteriori} correction to the electronic energy,
\begin{eqnarray}
\label{eq:edisp}
E_\text{disp}^\text{D3} \equiv  -\sum_{A,B>A} \ \left( s_6 \frac{C_6^{AB}}{R_{AB}^6}f^{(6)}_\text{damp}(R_{AB}) + s_8 \frac{C_8^{AB}}{R_{AB}^8}f^{(8)}_\text{damp}(R_{AB}) \right) \ ,
\end{eqnarray}
where $A$ and $B$ identify a pair of atoms in the system under consideration.
The dispersion coefficients $C_n^{AB}$ ($n = 6, 8$) are completely determined by the nuclear coordinates and atomic numbers and, therefore, not affected by the calibration procedure 
chosen for estimating optimal parameter values.
The damping function, $f^{(n)}_\text{damp}(R_{AB})$, is a function of  pairwise nuclear distances, $R_{AB}$, and requires further specification.
The original zero-damping function\cite{grimme2010},
\begin{equation}
f^{(n)}_\text{damp,zero}(R_{AB}) \equiv \frac{1}{1 + 6 \big(R_{AB} / (s_{r,n}R_0^{AB})\big)^{a_n}} \ ,
\end{equation}
contains one empirical (or adjustable) parameter, $s_{r,6}$, with fixed values for $s_{r,8}$, $a_6$, and $a_8$.
The improved BJ damping function\cite{grimme2011},
\begin{equation}
f^{(n)}_\text{damp,BJ}(R_{AB}) \equiv \frac{R_{AB}^n}{R_{AB}^n + (a_1R_0^{AB} + a_2)^n} \ ,
\end{equation}
contains two empirical parameters, $a_1$ and $a_2$.
In both cases, the cut-off radius $R_0^{AB}$ is defined as the square-root ratio of the corresponding $C_6^{AB}$ and $C_8^{AB}$ dispersion coefficients,
\begin{equation}
R_0^{AB} \equiv \sqrt{\frac{C_8^{AB}}{C_6^{AB}}} \ .
\end{equation}
Independent of the damping function chosen, the global scaling factor $s_8$ constitutes an additional empirical parameter, whereas $s_6$ is set to one.
The parameter $s_8$ is adjustable in order to compensate for double-counting effects\cite{chem_rev_2016_116_5105} and is functional-dependent (like the empirical 
parameters of the damping functions).
In line with the work by Grimme and coworkers\cite{chem_rev_2016_116_5105}, we define the dispersion coefficients $C_n^{AB}$ to be strictly positive, 
\textit{i.e.}, the calculated dispersion energy will be smaller than zero and 
asymptotically approach zero for increasing distances $R_{AB}$.

For all calculations of dispersion energies, $E_\text{disp}^\text{D3}$, Grimme's implementation of D3-zero and D3-BJ retrieved from the web page given in 
Ref.~\cite{dftd3_website} was employed. 
The program was minimally modified such that the values of the empirical parameters ($s_{r,6}$ and $s_8$ in the case of D3-zero, $a_1$, $s_8$, and $a_2$ in the case 
of D3-BJ) could be passed as command-line arguments instead of parsing an input file,
thereby leading to both significant computational speedup and easy parallelizability; features that are mandatory for efficient bootstrapping.

All reference molecular structures and dispersion energies were obtained from the following data sets of the GMTKN24 database\cite{j_chem_theory_comput_2010_6_107,gmtkn_website} 
for D3-zero, the GMTKN30 database \cite{j_chem_theory_comput_2011_7_291,gmtkn_website} for D3-BJ, and the supporting information of Ref.~\cite{grimme2010} for both D3-zero and D3-BJ:
S22, S22+, RG6, ADIM6, ACONF, CCONF, PCONF, and SCONF.
Note that the S22 and ACONF reference subsets are different in GMTKN24 and GMTKN30.
The reference dispersion energy is defined as the difference in the \textit{inter}molecular electronic interaction energy obtained with a (highly accurate) benchmark model 
and with a density functional (DF),
\begin{equation}
\Delta E_\text{disp}^\text{ref} \equiv	 E_\text{inter}^\text{benchmark} - E_\text{inter}^\text{DF} \ .
\end{equation}
The electronic interaction energy calculated with model $X$, in turn, is defined as the electronic energy difference between the entire \textit{system} and 
its \textit{subsystems} (here, two; ``sub1'' and ``sub2''),
\begin{equation}
E_\text{inter}^X \equiv E_\text{el,system}^X - \Big( E_\text{el,sub1}^X + E_\text{el,sub2}^X \Big) \ .
\end{equation}
Both $E_\text{inter}^\text{benchmark}$ and $E_\text{inter}^\text{DF}$ are taken from the databases mentioned above and were not recalculated by us. \textit{I.e.},
in this work, no electronic structure calculations had to be carried out.

All statistical protocols discussed in this paper were already introduced by us before\cite{proppe2017} and made available in our open-source calibration 
software \textsc{reBoot}\cite{proppe2017a}.
Here, we converted the Octave scripts of \textsc{reBoot} to the Python language.
These custom-made Python scripts rely on the packages \textsc{NumPy} 1.13.2\cite{comput_sci_eng_2011_13_22} and 
\textsc{SciPy} 0.19.1\cite{scipy}. 
For D3-zero, the calibration was accomplished by a brute-force search as implemented in \textsc{SciPy} 0.19.1, whereas for D3-BJ, differential 
evolution\cite{j_global_optim_197_11_341} (also implemented in \textsc{SciPy} 0.19.1) was chosen since the increased parameter space would render a brute-force 
search combined with bootstrapping inefficient. 
For D3-zero, we considered the following parameter ranges: $1.0 \leq s_{r,6} \leq 1.6$, $0.5 \leq s_8 \leq 1.1$, and for D3-BJ, it is 
$0.0 \leq a_1 \leq 0.7$, $0.0 \leq s_8 \leq 3.5$, $2.5 \leq a_2 \leq 6.5$. 
We carefully set all numerical convergence thresholds tight enough. 
Our preliminary tests showed that the default values yield satisfactory convergent behavior, except for the relative (dimensionless) \texttt{tol} parameter 
of the differential evolution algorithm, which was set to 0.001. 
For all bootstrap runs, a sample size of 10'000 was chosen (see also Section~\ref{sec:bootstrap}). 
For the generation of bootstrap samples, we exploited the default random number generator of Python as implemented in the module \texttt{random}.
The parameters obtained in the bootstrap runs will be made available as part of our program \textsc{BootD3} (see also Section~\ref{sec:conclusion}).

All plots were produced with \textsc{Mathematica} 11.0.1\cite{mathematica11_0_1_0}.

\section{Results and Discussion}
\label{sec:results}

\subsection{Parameter Optimization for D3-zero and D3-BJ}
\label{sec:params}

For our statistical analysis, we implemented a calibration routine which yields optimal values for the empirical parameters ($s_{r,6}$ and $s_8$ in case of D3-zero, 
and $a_1$, $s_8$, and $a_2$ in case of D3-BJ, respectively). 
This routine minimizes a cost function and forms the basis for our bootstrap analysis.
The cost function employed in this work is the mean absolute error (MAE) between the reference intermolecular dispersion energies, $\Delta E_\text{disp}^\text{ref}$ 
(taken from Refs.~\cite{j_chem_theory_comput_2010_6_107, gmtkn_website} for D3-zero and Refs.~\cite{j_chem_theory_comput_2011_7_291, gmtkn_website} for D3-BJ), and 
the intermolecular dispersion energies calculated for a given set of values of the empirical parameters, $\Delta E_\text{disp}^\text{D3}$,
\begin{equation}
\label{eq:mad}
\text{MAE} \equiv \frac{1}{N} \sum_{i=1}^N w_i \big\vert \Delta E_{\text{disp},i}^\text{ref} - \Delta E_{\text{disp},i}^\text{D3} \big\vert \ ,
\end{equation}
where
\begin{equation}
 \Delta E_{\text{disp},i}^\text{D3}  \equiv E_{\text{disp,system},i}^\text{D3} - \Big( E_{\text{disp,sub1},i}^\text{D3} + E_{\text{disp,sub2},i}^\text{D3} \Big) \ .
\end{equation}

Here, $N$ is the number of reference data points and $w_i$ refers to the weight of the $i$-th data point that will be discussed later.
Note that in the original work\cite{grimme2010, grimme2011} by Grimme and coworkers, the MAE is termed MAD (mean absolute deviation).
The MAD is a more general quantity which is defined as the average of the absolute deviations from a central point such as the mean of the absolute residuals (difference 
between reference and prediction) \cite{pernot2015}.
If that central point is set to zero, both MAD and MAE are identical quantities and as such employed in Grimme's and our studies.
We will apply the more specific term MAE in this paper, even though it is interchangeable with Grimme's definition of the MAD. 

First, we reproduced the results reported by Grimme and coworkers for D3-zero\cite{grimme2010}. 
In fact, when we adopted the parameter values reported by Grimme \textit{et al.}~to be optimal for D3-zero, we reproduced Grimme's MAEs for all density functionals studied
(PBE, revPBE, TPSS, BP86, B97-D, B3LYP, and B2PLYP). 
In a few cases, we observed negligible differences on the order of 0.01\,kcal\,mol$^{-1}$.
These results clearly suggest that our MAE definition provided in Eq.~(\ref{eq:mad}) is essentially identical to that used by Grimme and coworkers.
Two alternative MAE definitions,
\begin{equation}
\text{MAE2} \equiv \frac{1}{N} \sum_{i=1}^N \big\vert \Delta E_{\text{disp},i}^\text{ref} - \Delta E_{\text{disp},i}^\text{D3} \big\vert
\end{equation}
and
\begin{equation}
\text{MAE3} \equiv \frac{1}{\sum_{i=1}^N w_i} \sum_{i=1}^N w_i \big\vert \Delta E_{\text{disp},i}^\text{ref} - \Delta E_{\text{disp},i}^\text{D3} \big\vert \ ,
\end{equation}
resulted in significant deviations from the MAEs reported by Grimme and coworkers.

Concerning the MAEs reported by Grimme \textit{et al.\@} for D3-BJ\cite{grimme2011}, we found that we could reproduce these values for the S22 and PCONF subsets for all 
density functionals studied. 
However, when comparing the MAEs over the entire reference set, we found significant deviations for the density functionals BP86 (0.16\,kcal\,mol$^{-1}$), B97-D 
(0.03\,kcal\,mol$^{-1}$), B3LYP (0.04\,kcal\,mol$^{-1}$), and B2PLYP (0.05\,kcal\,mol$^{-1}$).
In all these cases, our MAEs were higher than those reported in Ref.~\cite{grimme2011} (note that the MAEs in that reference had to be extracted from a 
figure).
We may assume that these deviations result from differences in the other subsets; especially ACONF, since this reference data set is, as S22, different 
in GMTKN24 and GMTKN30\cite{gmtkn_website}.
However, subset-specific MAEs were only reported for S22 and PCONF in Ref.~\cite{grimme2011}.

\begin{figure}[H]
\begin{center}
\includegraphics[width=\textwidth]{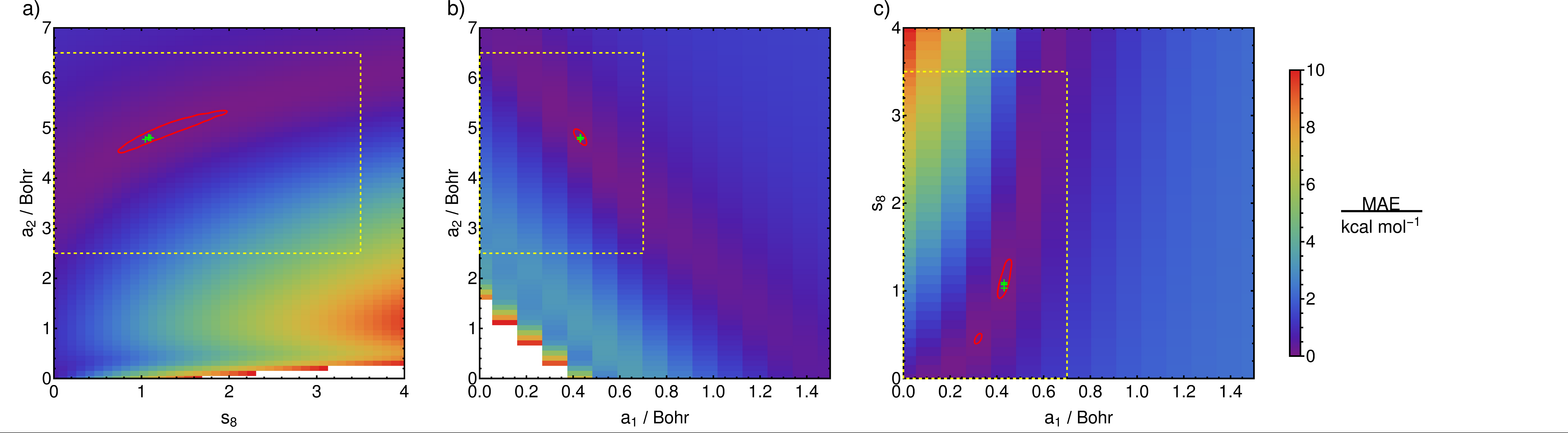}
\end{center}
\caption{\label{fig:mad}\small 
MAE of Eq.~(\ref{eq:mad}) for PBE-D3-BJ dispersion energies from reference dispersion energies (reference energies taken from Ref.~\cite{grimme2011}) in kcal\,mol$^{-1}$
as a function of the three empirical 
parameters $a_1$, $s_8$, and $a_2$. 
The yellow dashed rectangle specifies the bounds of the differential evolution optimization. 
The red line refers to an MAE of 0.5\,kcal\,mol$^{-1}$.
The crosses show the results from five individual optimizations by means of differential evolution. 
To produce this figure, in a), $a_1$ was fixed to a value of 0.4309\,Bohr; in b), $s_8$ was set to 1.0892; in c), $a_2$ was fixed to 4.8327\,Bohr}
\end{figure}

It is instructive to visually inspect the cost function to be minimized. The cost function for the PBE density functional is shown in Fig.~\ref{fig:mad}.
The MAE varies between zero and very large values of more than 10\,kcal\,mol$^{-1}$ over the large parameter range considered. 
However, the optimization problem is not too complex since the parameter dependence of the cost function does not exhibit a rugged form, but is usually convex.
Still, there can be cases where the cost function exhibits more than a single minimum, as shown in Fig.~\ref{fig:mad}~c). In this case, the cost function has two minima, 
one around $s_8 \approx 0.4$ and a second one around $s_8 \approx 1.1$. 
A more detailed investigation shows that the minimum around $s_8 \approx 1.1$ is the global minimum. 
As indicated by the crosses in Fig.~\ref{fig:mad}, the differential evolution algorithm adopted in our study reliably finds the global minimum, also for nonconvex cost 
functions.

The optimal parameters obtained with our optimization procedure are reported in Tables~\ref{tab:param-zero}~and~\ref{tab:param-bj} for D3-zero and D3-BJ, respectively.
At this point, we should note that we report parameters with a precision of four decimal places in this work (in some exceptional and justified cases we will 
divert from this choice). 
We emphasize that this choice does not indicate the number of \textit{statistically} significant figures; we chose this information content to be consistent 
with the representation of the empirical D3-BJ parameters in Grimme's D3 implementation\cite{dftd3_website}.
In fact, as we will see, (statistical) parameter uncertainty usually already affects the third decimal place. Still, we will report more digits for the 
sake of consistency and in order to rule out any rounding issues. Of course, many more digits were taken into account in the calculations, so that numerical
(cut-off) errors have a negligible effect on our results.

\begin{table}[H]
\renewcommand{\baselinestretch}{1.0}
\renewcommand{\arraystretch}{1.0}
\caption{\label{tab:param-zero}\small 
Optimized D3-zero parameters $s_{r,6}$ and $s_8$ for all density functionals studied in this work.
The original parameters reported by Grimme \textit{et al.}\cite{grimme2010} are given in parentheses (note that only three decimal places are defined in Grimme's D3 
implementation\cite{dftd3_website}). 
In both cases, the same reference set \cite{grimme2010} was employed.}
\begin{center}
\begin{tabular}{l r r r r} \hline \hline
Functional & $s_{r,6}$ / Bohr & $s_8$    & MAE / kcal\,mol\textsuperscript{$-$1} \\
\hline 
PBE        & 1.2794 (1.217)   & 0.7798 (0.722)  & 0.52 (0.56) \\
revPBE     & 0.9075 (0.923)   & 0.9514 (1.010)  & 0.37 (0.37) \\
TPSS       & 1.1768 (1.166)   & 1.0962 (1.105)  & 0.45 (0.48) \\ 
BP86       & 1.1914 (1.139)   & 1.6878 (1.683)  & 0.56 (0.65) \\
B97-D      & 0.8830 (0.892)   & 0.8666 (0.909)  & 0.38 (0.37) \\
B3LYP      & 1.2665 (1.261)   & 1.6485 (1.703)  & 0.38 (0.37) \\
B2PLYP     & 1.3012 (1.332)   & 0.9458 (1.000)  & 0.23 (0.21) \\
\hline
\hline
\end{tabular}
\renewcommand{\baselinestretch}{1.0}
\renewcommand{\arraystretch}{1.0}
\end{center}
\end{table}

We obtained slightly different parameter values for D3-zero compared to those reported by Grimme and coworkers.
However, all qualitative trends between the Grimme parameterization and ours are the same.
For example, in both parameterizations, the largest value for $s_{r,6}$ was found for B2PLYP, while the smallest value was found for B97-D. 
The differences between the MAEs obtained with the two parameterizations were also very small. 
Still, the two parameterizations sometimes led to significant MAE differences, especially for BP86, where we found an MAE of 0.56\,kcal\,mol$^{-1}$ while the original 
D3-zero parameterization produced an MAE of 0.65\,kcal\,mol$^{-1}$.
These differences for D3-zero are most likely a result of different cost functions employed for parameter optimization.
Grimme and coworkers briefly mention the use of a weighted least-squares calibration procedure \cite{grimme2010, grimme2011}.
However, when we substituted the absolute value of residuals, $\vert E_{\text{disp},i}^\text{ref} - E_{\text{disp},i}^\text{D3} \vert$, in Eq.~(\ref{eq:mad}) by its 
squared value, $\big( E_{\text{disp},i}^\text{ref} - E_{\text{disp},i}^\text{D3} \big)^2$, the deviations from Grimme's parameters were still significant and even 
increased in some cases.
In the following, all results reported here refer to the MAE cost function introduced in Eq.~(\ref{eq:mad}).
We emphasize that the representation of residuals (here, absolute versus squared values) in the cost function does not affect the conclusions drawn in 
this paper.

\begin{table}[h]
\renewcommand{\baselinestretch}{1.0}
\renewcommand{\arraystretch}{1.0}
\caption{\label{tab:param-bj}\small 
Optimized D3-BJ parameters $a_1$, $s_8$, and $a_2$ for all density functionals studied in this work.
The original parameters reported by Grimme \textit{et al.}\cite{grimme2011} are given in parentheses. In both cases, the same reference set\cite{grimme2011} was employed. 
Note that the MAE values from Ref.~\cite{grimme2011} had to be extracted from a graphical figure.}
\begin{center}
\begin{tabular}{l r r r r} \hline \hline
Functional & $a_1$ / Bohr     & $s_8$     & $a_2$ / Bohr     & MAE / kcal\,mol\textsuperscript{$-$1} \\
\hline 
PBE        & 0.4309 (0.4289)  & 1.0892 (0.7875)  & 4.8327 (4.4407)  & 0.49 (0.50)  \\
revPBE     & 0.4654 (0.5238)  & 1.7394 (2.3550)  & 3.4591 (3.5016)  & 0.39 (0.42) \\
TPSS       & 0.4557 (0.4535)  & 2.0871 (1.9435)  & 4.5886 (4.4752)  & 0.44 (0.46) \\
BP86       & 0.4245 (0.3946)  & 3.3975 (3.2822)  & 4.8921 (4.8516)  & 0.54 (0.48) \\
B97-D      & 0.5059 (0.5545)  & 1.5949 (2.2609)  & 3.1014 (3.2297)  & 0.36 (0.40) \\
B3LYP      & 0.3876 (0.3981)  & 1.8660 (1.9889)  & 4.4814 (4.4211)  & 0.28 (0.27)  \\
B2PLYP     & 0.3371 (0.3451)  & 0.9518 (1.0860)  & 4.7394 (4.7735)  & 0.20 (0.17)  \\
\hline
\hline
\end{tabular}
\renewcommand{\baselinestretch}{1.0}
\renewcommand{\arraystretch}{1.0}
\end{center}
\end{table}

For D3-BJ, we find more pronounced deviations from the parameters reported by Grimme and coworkers\cite{grimme2011} compared to D3-zero in some cases, \textit{e.g.}, for 
$s_8$ when considering PBE and B97-D (see Table~\ref{tab:param-bj}).
Due to the additional empirical parameter to be optimized in D3-BJ compared to D3-zero, the solution space is of higher dimension and may, therefore, be more sensitive 
to differences in the calibration procedure (\textit{e.g.}, cost function, reference set, optimization algorithm).
Still, as in the case of D3-zero, all qualitative trends are the same.
We should note that even though the differences between the MAEs are only on the order of a few hundredths kcal\,mol$^{-1}$, their relative differences can be significant.
For example, for both BP86 and B97-D the MAE obtained with our parameterization is lowered by 10\,\% compared to the MAE evaluated with the original parameterization, 
whereas for B2PLYP, the MAE obtained with the original parameterization is lower by 15\,\%. This last result further corroborates our assumption
mentioned above in the discussion of different MAE definitions that the 
reference set employed in this study for D3-BJ is different from the one used in the original parameterization of 
D3-BJ\cite{grimme2011}.
Since we optimized the D3-BJ parameters by minimizing the MAE, another parameterization cannot lead to a lower MAE.
In fact, when adopting the original D3-BJ parameters, we consistently find higher MAEs compared to our parameterization.
However, as can be seen in Table~\ref{tab:param-bj}, Grimme \textit{et al.\@} report lower MAEs for BP86, B3LYP, and B2PLYP compared to our data.

It is imperative to test the performance of a given set of empirical parameters on an independent validation set, \textit{i.e.}, a data set which has not been employed 
in the calibration routine. 
To this end, we provide the MAEs resulting from our D3-BJ parameterizations for three validation sets\,---\,all taken from the 
GMTKN30 database\cite{j_chem_theory_comput_2011_7_291}\,---\,in Table~\ref{tab:validation}.

\begin{table}[H]
\renewcommand{\baselinestretch}{1.0}
\renewcommand{\arraystretch}{1.0}
\caption{\label{tab:validation}\small 
MAEs according to Eq.~(\ref{eq:mad}) for three different validation sets taken from the GMTKN30 database\cite{j_chem_theory_comput_2011_7_291} resulting 
from our D3-BJ parameterization. 
The values in parentheses were obtained by us with the original D3-BJ parameterization by Grimme and coworkers\cite{grimme2011}. 
All data are given in kcal\,mol\textsuperscript{$-$1}.}
\begin{center}
\begin{tabular}{l r r r r} \hline \hline
Functional & HEAVY28       & AL2X         & DARC  \\
\hline 
PBE        &  0.32 (0.36)  & 2.14 (2.35)  & 4.02 (3.62)  \\
revPBE     &  0.33 (0.26)  & 2.89 (2.34)  & 3.30 (3.74)  \\
TPSS       &  0.29 (0.31)  & 3.38 (3.55)  & 5.48 (5.18)  \\
BP86       &  0.33 (0.55)  & 2.44 (3.31)  & 4.50 (3.82)  \\
B97-D      &  0.31 (0.28)  & 3.10 (3.77)  & 9.46 (10.41) \\
B3LYP      &  0.24 (0.28)  & 1.79 (1.62)  & 7.99 (7.69)  \\
B2PLYP     &  0.19 (0.23)  & 1.07 (1.07)  & 4.05 (4.00)  \\
\hline
\hline
\end{tabular}
\renewcommand{\baselinestretch}{1.0}
\renewcommand{\arraystretch}{1.0}
\end{center}
\end{table}

\begin{table}[h]
\renewcommand{\baselinestretch}{1.0}
\renewcommand{\arraystretch}{1.0}
\caption{\label{tab:al2x_errors}\small 
Absolute values of residuals, $e_i = \big\vert\Delta E_{\text{disp},i}^\text{ref} - \Delta E_{\text{disp},i}^\text{D3}\big\vert $, in kcal\,mol$^{-1}$, for the AL2X validation set obtained with the 
original D3-BJ parameterization by Grimme and coworkers\cite{grimme2011}, $e_{\text{org},i}$, and the one introduced in this paper, $e_{\text{new},i}$.
}
\begin{center}
\begin{tabular}{l r r r} \hline \hline
System & $e_{\text{org},i}$ & $e_{\text{new},i}$ & $e_{\text{org},i} / e_{\text{new},i}$ \\
\hline 
Al$_2$H$_6$ & 3.51 & 3.04 & 115.4\,\% \\
Al$_2$F$_6$ & 6.58 & 7.11  & 92.5\,\% \\
Al$_2$Cl$_6$ & 2.31 & 0.93 & 246.9\,\% \\
Al$_2$Br$_6$ & 4.51 & 2.67 & 168.5\,\% \\
Al$_2$Me$_4$ & 1.35 & 0.68 & 196.9\,\% \\
Al$_2$Me$_5$ & 2.04 & 1.07 & 190.7\,\% \\
Al$_2$Me$_6$ & 2.87 & 1.57 & 183.4\,\% \\
\hline
\hline
\end{tabular}
\renewcommand{\baselinestretch}{1.0}
\renewcommand{\arraystretch}{1.0}
\end{center}
\end{table}

The reliability of our new parameterization is comparable to that of the original parameterization.
The difference between the corresponding MAEs for the two parameterizations is usually very small (on the order of 0.01\,kcal\,mol$^{-1}$). 
In a few cases (\textit{e.g.}, AL2X with BP86 and DARC with B97-D), the difference between the MAEs resulting from the two parameterizations is significantly larger, 
reaching almost 1\,kcal\,mol$^{-1}$.
If the overall effect of the two slightly different parameterizations may become large in some cases, this will raise the question whether such effects based on 
a specific choice for a parameter set are actually negligible in routine quantum chemical calculations.

\begin{table}[h]
\renewcommand{\baselinestretch}{1.0}
\renewcommand{\arraystretch}{1.0}
\caption{\label{tab:darc_errors}\small 
Absolute values of residuals, $e_i = \big\vert \Delta E_{\text{disp},i}^\text{ref} - \Delta E_{\text{disp},i}^\text{D3} \big\vert $, in kcal\,mol$^{-1}$, for the DARC validation set obtained with the 
original D3-BJ parameterization by Grimme and coworkers\cite{grimme2011}, $e_{\text{org},i}$, and the one introduced in this paper, $e_{\text{new},i}$.
}
\begin{center}
\begin{tabular}{l r r r} \hline \hline
System & $e_{\text{org},i}$ & $e_{\text{new},i}$ & $e_{\text{org},i} / e_{\text{new},i}$ \\
\hline 
P1 & 6.74 &  7.67 & 113.8\,\% \\
P2 & 2.81 & 3.90 & 138.6\,\% \\
P3 & 10.06 & 11.07 & 109.9\,\% \\
P4 & 6.17 & 7.34 & 119.0\,\% \\
P5 & 8.63 & 9.61 & 111.4\,\% \\
P6 & 5.12 & 6.35 & 124.1\,\% \\
P7 & 11.93 & 12.77 & 107.0\,\% \\
P7X & 12.36 & 13.24 & 107.1\,\% \\
P8 & 11.90 & 12.71 & 106.8\,\% \\
P8X & 12.23 & 13.09 & 107.1\,\% \\
P9 & 11.17 & 12.02 & 107.6\,\% \\
P9X & 11.21 & 12.12 & 108.1\,\% \\
P10 & 11.08 & 11.90 & 107.4\,\% \\
P10X & 11.01 & 11.89 & 108.0\,\% \\
\hline
\hline
\end{tabular}
\renewcommand{\baselinestretch}{1.0}
\renewcommand{\arraystretch}{1.0}
\end{center}
\end{table}

Note that in the case of AL2X (with BP86) and DARC (with B97-D), also the absolute errors are comparatively large. 
To investigate the reasons for this, we calculated the errors for the individual data points of these two sets.
On average, the MAE resulting from Grimme's original parameterization was 135.6\,\% of the MAE resulting from our parameter values in the case of the AL2X data set 
(Table~\ref{tab:al2x_errors}). 
This deviation was mostly created by Al$_2$Cl$_6$, but also by Al$_2$Me$_4$, Al$_2$Me$_5$, and Al$_2$Me$_6$.
When closer analyzing the dispersion-energy difference determined for the Al$_2$Cl$_6$ data point, we found that for AlCl$_3$ (one of the two identical monomers of 
Al$_2$Cl$_6$), the total dispersion energy is $-$5.87\,kcal\,mol$^{-1}$ for Grimme's original parameterization and $-$4.86\,kcal\,mol$^{-1}$ for our new parameters.  
When analyzing the individual pairs, we found that for all atom pairs, the dispersion energy is smaller in Grimme's case. 
We made the same observation for Al$_2$Cl$_6$. 
For the DARC validation set, we found the same trend, but the relative deviations are less pronounced (Table~\ref{tab:darc_errors}).

We suggest two reasons for large absolute dispersion-energy errors of individual systems.
First, the prediction model (here, D3-BJ) is inadequate such that certain domains of the chemical space are systematically misdescribed, no matter what parameterization 
would be chosen (see also Section~\ref{sec:bootstrap} below).
Second, since the number of pairwise interactions depends on system size, small inaccuracies in the calibration procedure may accumulate to large errors for large 
molecular systems.
This latter error type will be discussed in detail in Section~\ref{sec:size}.

\subsection{Consequences of Different Weights on Model Performance}
\label{sec:weighting}

In the original calibration routine of Grimme and coworkers\cite{grimme2010}, the RG6 set (containing six rare gas dimers) was given a weight of 20, while all 
other data points were assigned a weight of one. 
It is not immediately clear why such a weighting scheme should be  adopted, nor is this choice explained in Refs.~\cite{grimme2010, grimme2011}. 
Therefore, we decided to carry out a reparameterization where this increased weighting was removed, hence considering all data points on an equal footing. 
In the following, we refer to this variant as the ``uniform weighting scheme'', for which we find MAE $=$ MAE2 $=$ MAE3.
The resulting optimal parameter values together with their corresponding MAEs with respect to the reference set are listed in Table~\ref{tab:param-weighting}.

By comparison with Table~\ref{tab:param-bj}, we note that the parameter values sometimes change significantly. 
For example, in the case of PBE, $a_1$ decreases from 0.4309\,Bohr~to 0.0181\,Bohr~when adopting the uniform weighting scheme. 
Also for TPSS, B97-D, and BP86 we note rather strong differences between the parameter values obtained with the original and the uniform weighting scheme.
For the other density functionals, however, the changes are much smaller. 
It is not surprising that the MAE with respect to the reference set is generally smaller in the case of the uniform weighting scheme.
Increasing the weight of a given data point can be achieved by repeatedly adding the data point to the reference set until the desired weight has been reached;
\textit{cf.\@} Eq.~(\ref{eq:mad}). 
For example, in the present case, each data point of the RG6 set would be added twenty times to the reference set. 
In the uniform weighting scheme, each of these data points occurs only once, and so the reference set is effectively smaller.
It is natural that the (cumulative) error with respect to a reference set always depends on its size. 
Therefore, we need to analyze the MAE obtained on some validation sets in order to develop a clear picture of the reliability of this new parameterization.

\begin{table}[H]
\renewcommand{\baselinestretch}{1.0}
\renewcommand{\arraystretch}{1.0}
\caption{\label{tab:param-weighting}\small 
Optimized D3-BJ parameters $a_1$, $s_8$, and $a_2$ for all density functionals studied in this work when a uniform weighting scheme was employed.}
\begin{center}
\begin{tabular}{l r r r r} \hline \hline
Functional & $a_1$ / Bohr  & $s_8$  & $a_2$ / Bohr  & MAE / kcal\,mol\textsuperscript{$-$1} \\
\hline 
PBE        & 0.0181        & 0.8522        & 6.4974        &   0.43 \\  
revPBE     & 0.4632        & 1.7402        & 3.4677        &   0.37 \\
TPSS       & 0.4228        & 2.0633        & 4.7281        &   0.41 \\
BP86       & 0.4910        & 3.4927        & 4.6054        &   0.41 \\
B97-D      & 0.3972        & 1.5116        & 3.5264        &   0.33 \\
B3LYP      & 0.3289        & 1.8484        & 4.7367        &   0.25 \\
B2PLYP     & 0.3355        & 0.9462        & 4.7423        &   0.17 \\
\hline
\hline
\end{tabular}
\renewcommand{\baselinestretch}{1.0}
\renewcommand{\arraystretch}{1.0}
\end{center}
\end{table}

\begin{table}[H]
\renewcommand{\baselinestretch}{1.0}
\renewcommand{\arraystretch}{1.0}
\caption{\label{tab:validation-weighting}\small MAEs  in kcal\,mol\textsuperscript{$-$1} for three different validation sets taken from the GMTKN30 database\cite{j_chem_theory_comput_2011_7_291}
resulting from our D3-BJ parameterizations when adopting a uniform weighting scheme.}
\begin{center}
\begin{tabular}{l r r r r} \hline \hline
Functional & HEAVY28 & AL2X  & DARC  \\
\hline 
PBE        & 1.07    & 5.54  & 3.62 \\
revPBE     & 0.34    & 2.92  & 3.29 \\
TPSS       & 0.34    & 3.60  & 5.43 \\
BP86       & 0.25    & 1.92  & 4.55 \\
B97-D      & 0.64    & 2.06  & 8.74 \\
B3LYP      & 0.42    & 1.41  & 7.84 \\
B2PLYP     & 0.20    & 1.07  & 4.05 \\
\hline
\hline
\end{tabular}
\renewcommand{\baselinestretch}{1.0}
\renewcommand{\arraystretch}{1.0}
\end{center}
\end{table}

The MAEs obtained for the validation sets HEAVY28, AL2X, and DARC are provided in Table~\ref{tab:validation-weighting}.
The uniform weighting scheme sometimes leads to a significantly reduced reliability (\textit{cf.\@} Table~\ref{tab:validation}). 
For instance, for PBE, the MAE (Eq.~(\ref{eq:mad})) with respect to the AL2X set increases from 2.14\,kcal\,mol$^{-1}$ to 5.54\,kcal\,mol$^{-1}$. 
In some cases, the uniform weighting scheme also leads to improved results. 
Most interestingly, for the DARC set, a lower MAE is found for most density functionals studied in this work.

Therefore, neither the original weighting scheme adopted by Grimme and coworkers nor the uniform weighting scheme is optimal in all cases, but 
both weighting schemes lead to a good overall reliability.
From the results above it can be expected that the numerical effect that different weighting schemes have on dispersion corrections in actual quantum chemical 
calculations can become significant. 
We will show in Section~\ref{sec:size} that different weighting schemes may have a dramatic impact on the dispersion energy error for large molecular systems.

\subsection{Bootstrapping Model Prediction Uncertainty}
\label{sec:bootstrap}

The MPU is an important key characteristic of every model. 
However, it is often difficult to quantify reliably the MPU. 
A first estimate for the MPU is the residual error with respect to the reference set (such as the MAE), but such an estimate depends on the actual reference set chosen. 
To obtain an MPU measure independent of the reference set, one often resorts to validation sets (as we have done above in Sections~\ref{sec:params} and \ref{sec:weighting}).

Another way of estimating the MPU without the need of additional validation sets is the so-called nonparametric bootstrapping approach\cite{ann_statist_1979_7_1}, which 
we recently implemented for the calibration of physicochemical property models and discussed in detail at the example of $^{57}$Fe M\"ossbauer isomer shift 
prediction\cite{proppe2017}.
In this approach, one generates a large collection of data sets, sampled from the original reference set, to reproduce approximately the population distribution 
underlying the original reference set.
For each of these data sets (bootstrap samples), an individual calibration is carried out, leading to an ensemble of parameter sets.
This ensemble allows us obtain an estimate $m$ for the expected value $\mu$, 
\begin{equation}
\mu(v) \approx	m(v) \equiv \frac{1}{B} \sum_{b=1}^B v_b \ ,
\end{equation}		
as well as an estimate $s^2$ for the (co)variances $\sigma^2$,
\begin{equation}
\sigma^2(v,w) \approx s^2(v,w) \equiv \ \frac{1}{B-M} \sum_{b=1}^B \big(v_b - m(v)\big) \big(w_b - m(w)\big) ,
\end{equation}		
 of the generic parameters $v$ and $w$.
 Here, $B$ and $M$ denote the numbers of bootstrap samples and empirical parameters, respectively, and $v_b$ refers to the optimal $v$-value for the $b$-th bootstrap sample.
 For $v = w$, we additionally define $s^2(v,v) \equiv s^2(v)$.
 Moreover, instead of studying the low-order moments (expected value, covariance) of the parameter distributions, their specific functional form can be approximated 
through histograms that, \textit{e.g.}, facilitate to judge whether the original reference set \cite{grimme2011} is balanced or skewed (as will be discussed below).
Furthermore, the ensemble of bootstrap samples allows us to generate an ensemble of MAEs.
The mean over all MAEs is an estimate to the actual MPU,
\begin{equation}
\label{eq:mpu}
\text{MPU} \approx m(\text{MAE}) = \frac{1}{B} \sum_{b=1}^B \text{MAE}_b \ ,
\end{equation}
where $\text{MAE}_b$ constitutes the MAE \textit{determined from} the $b$-th parameter set \textit{with respect to} the original reference set.
The difference between max(MAE) and min(MAE) defines the maximum MAE spread sampled.

\begin{table}[h]
\renewcommand{\baselinestretch}{1.0}
\renewcommand{\arraystretch}{1.0}
\caption{\label{tab:param-bootstrap}\small Mean ($m$) and standard deviation ($s$) of the empirical D3-BJ parameters $a_1$, $s_8$, and $a_2$ as obtained
from bootstrapping ($B =$ 10'000) for all density functionals studied in this work. All values for $m(a_1)$, $s(a_1)$, $m(a_2)$, and $s(a_2)$
are given in Bohr.}
\begin{center}
\begin{tabular}{l r r r r r r} \hline \hline
Functional & $m(a_1)$        & $s(a_1)$         &  $m(s_8)$          & $s(s_8)$          &  $m(a_2)$         & $s(a_2)$ \\
\hline 
PBE        & 0.4191         & 0.0716            & 1.2367          & 0.3782            & 4.9545          & 0.2886   \\
revPBE     & 0.4764         & 0.0319            & 1.7072          & 0.1258            & 3.3687          & 0.1518   \\
TPSS       & 0.4404         & 0.0423            & 2.1088          & 0.3109            & 4.6705          & 0.2242   \\
BP86       & 0.4177         & 0.0520            & 3.3954          & 0.1159            & 4.9232          & 0.2486   \\
B97-D      & 0.5078         & 0.0597            & 1.5952          & 0.1131            & 3.0913          & 0.2429   \\
B3LYP      & 0.3800         & 0.0423            & 1.8781          & 0.1430            & 4.5183          & 0.2086   \\
B2PLYP     & 0.3197         & 0.0736            & 0.9513          & 0.0912            & 4.8235          & 0.3519   \\
\hline
\hline
\end{tabular}
\renewcommand{\baselinestretch}{1.0}
\renewcommand{\arraystretch}{1.0}
\end{center}
\end{table}

Table~\ref{tab:param-bootstrap} contains the mean values for all three empirical D3-BJ parameters as obtained from our bootstrapping procedure together with the 
corresponding standard deviations. 
By comparison with Table~\ref{tab:param-bj}, we see that the parameters obtained from bootstrapping are not significantly different from the original parameters, 
and all qualitative trends are the same.
Concerning the standard deviations given in Table~\ref{tab:param-bootstrap}, we should note that these are the direct standard deviations of the parameter distributions 
as obtained from the bootstrapping procedure. 
Usually, however, one is not so much interested in this standard deviation but more in the standard deviation of \textit{the mean} of these distributions. 
Assuming these parameters to be distributed normally, the standard deviation $s\big(m(v)\big)$ of the mean $m(v)$ of a given parameter $v$ can be determined on the basis 
of the well-known formula $s\big(m(v)\big) = s(v)/\sqrt{B}$, where $B$, in our case, equals 10'000. 
Therefore, the standard deviation of the mean of the empirical parameters $a_1$, $s_8$, and $a_2$ is smaller by a factor of 100 compared to the direct standard deviations. 
Hence, based on the results shown in Table~\ref{tab:param-bootstrap} and assuming a Gaussian distribution, we found the following empirical parameters 
together with a two-sided 95\,\% confidence interval when employing the PBE density functional: $a_1$ = 0.4191\,Bohr~$\pm$ 0.0014\,Bohr; $s_8$ = 1.237~$\pm$ 0.007; 
$a_2$ = 4.955\,Bohr~$\pm$ 0.006\,Bohr
To assess the reliability of assuming a Gaussian distribution, we also bootstrapped the mean of the PBE-D3-BJ parameters.
We found the following mean values and two-sided 95\,\% confidence intervals: $a_1$ = 0.4191\,Bohr~$\pm$ 0.0014\,Bohr; $s_8$ = 1.237~$\pm$ 0.007; 
$a_2$ = 4.955\,Bohr~$\pm$ 0.006\,Bohr
These results are literally identical to the ones based on the assumption of a normal distribution and we may safely assume that this
assumption is also true for all other density functionals.
Strictly speaking, it is not reasonable to report more than three decimal places for $s_8$ and $a_2$, since any figure after the third decimal place is affected 
by parameter uncertainty.
However, as we have already stated above, we will continue to report four decimal places for the sake of consistency (also recall that many more digits were
taken into account during the calculations so that cut-off errors can be neglected).

We now analyze the mean prediction errors obtained from the individual bootstrap samples. 
A summary is given in Table~\ref{tab:bootstrap-errors}. 
The average MAEs obtained from bootstrapping are only slightly larger than the MAEs obtained with the original parameterization (\textit{cf.\@} Table~\ref{tab:param-bj}), 
even though the largest MAEs are often more than twice as large as the mean.

\begin{table}[H]
\renewcommand{\baselinestretch}{1.0}
\renewcommand{\arraystretch}{1.0}
\caption{\label{tab:bootstrap-errors}\small 
min(MAE), max(MAE) and $m$(MAE), \textit{cf.\@} Eq.~(\ref{eq:mpu}), as obtained from bootstrapping for all density functionals studied in this work. 
All data are given in kcal\,mol\textsuperscript{$-$1}.}
\begin{center}
\begin{tabular}{l r r r} \hline \hline
Functional & min(MAE) & max(MAE) & $m$(MAE) \\
\hline 
PBE        & 0.49         & 0.80        & 0.50       \\    
revPBE     & 0.39         & 0.80        & 0.40       \\
TPSS       & 0.44         & 0.93        & 0.45       \\             
BP86       & 0.54         & 1.16        & 0.56       \\         
B97-D      & 0.36         & 0.67        & 0.38       \\
B3LYP      & 0.28         & 0.72        & 0.29       \\
B2PLYP     & 0.20         & 0.43        & 0.21       \\
\hline
\hline
\end{tabular}
\renewcommand{\baselinestretch}{1.0}
\renewcommand{\arraystretch}{1.0}
\end{center}
\end{table}

\begin{figure}[H]
\begin{center}
\includegraphics[width=\textwidth]{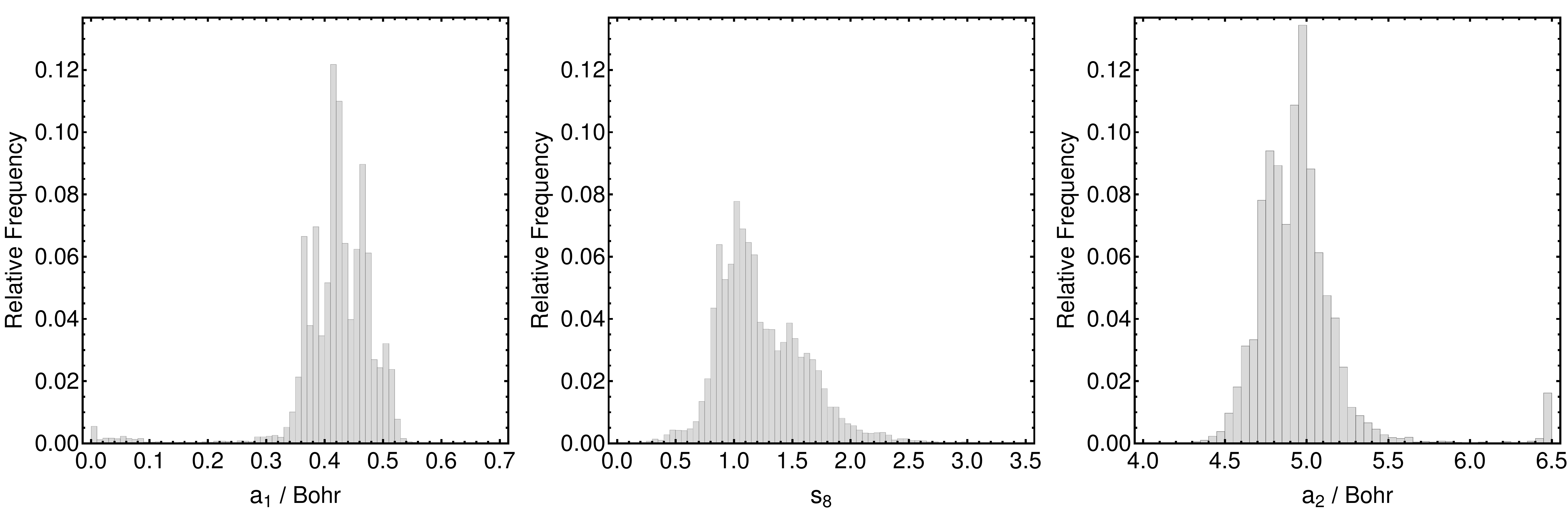}
\end{center}
\caption{\label{fig:bootstrap}\small Histograms of the PBE-D3-BJ parameters $a_1$, $s_8$, and $a_2$ as obtained from 10'000 bootstrap samples.}
\end{figure}

Histograms of the distributions of the results obtained for $a_1$, $s_8$, and $a_2$ are presented in Fig.~\ref{fig:bootstrap}.
While most parameters are centered around the optimal value obtained by the standard optimization procedure as discussed in Section~\ref{sec:params}, we note some distinct 
outliers for all parameters, \textit{i.e.}, there are a few bootstrap samples which lead to very different parameters. 
This finding suggests that the original reference set is not balanced but skewed in the sense that it contains a few data points which significantly bias the optimization.

\begin{figure}[h]
\begin{center}
\includegraphics[width=\textwidth]{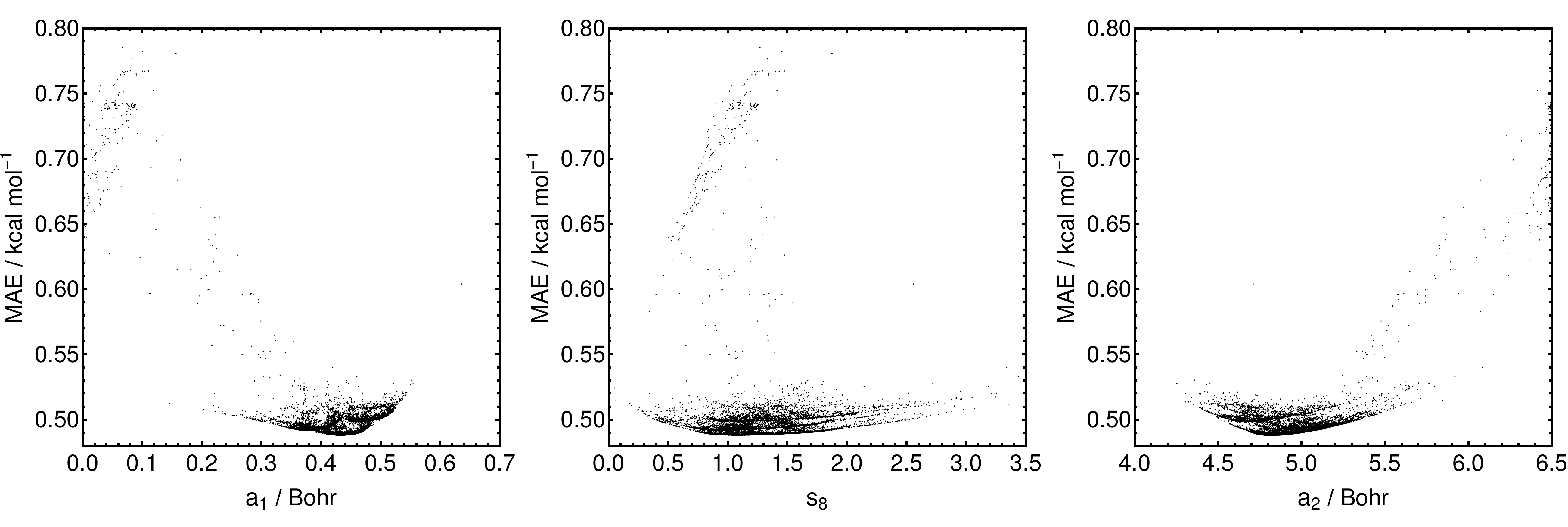}
\end{center}
\caption{\label{fig:bootstrap_detail}\small 
The MAE as a function of the PBE-D3-BJ parameters $a_1$, $s_8$, and $a_2$ obtained from 10'000 bootstrap samples. Each of these samples is represented by one dot 
in each of the three panels.
}
\end{figure}

To examine this issue in more detail, we study the MAE with respect to the original reference set as a function of the empirical D3-BJ parameters $a_1$, $s_8$, and 
$a_2$ for all 10'000 bootstrap samples (Fig.~\ref{fig:bootstrap_detail}).
We observe an interesting pattern in each MAE--parameter plot.
It appears that subsets of MAE--parameter pairs form parabolic clusters.
When analyzing all MAEs smaller than 0.49\,kcal\,mol$^{-1}$, we cannot find a single molecular system which is either present or absent in all data points.
Moreover, we do not find that any reference subset is over- or underrepresented (on average) in the bootstrap samples yielding an MAE $<$0.49\,kcal\,mol$^{-1}$.
On the contrary, we find that every bootstrap sample that yields an MAE larger than 0.70\,kcal\,mol$^{-1}$ does not contain any of the following three RG6 systems: 
the Rn--Rn dimer, the Rn--Xe dimer, and the Xe--Xe dimer.
It is not straightforward to decide whether the absence of these dimers biases the parameterization or not.
Intuitively, one may argue that this is not the case since each of these dimers is assigned a weight of 20.
Consequently, any contribution of these dimers to the MAE is also given this weight, \textit{cf.}~Eq.~\ref{eq:mad}.

We discuss the issue of biasing reference data by means of the jackknife method \cite{ann_statist_1979_7_1, riu2003} in the next Section.

\subsection{Jackknifing the Bias of Reference Data}
\label{sec:jackknife}

In the first step of the jackknife method, one generates $N$ new data sets (jackknife samples) from the original reference set containing $N$ data points, each 
containing all but one data point from the original reference set. 
In each of these jackknife samples, the missing data point is a different one; hence, in the first jackknife sample, the first data point is absent, whereas in the 
second sample, the second point is absent, and so forth. 
Then, the optimal parameter values are determined for each of these jackknife samples. 
If the original reference set is not biased, the distribution of the parameter values obtained by the jackknife method should indicate a smooth function. 
If, however, the original reference set is biased by certain data points, the jackknife samples with those data points removed will indicate nonsmooth parameter distributions.

\begin{figure}[H]
\begin{center}
\includegraphics[width=\textwidth]{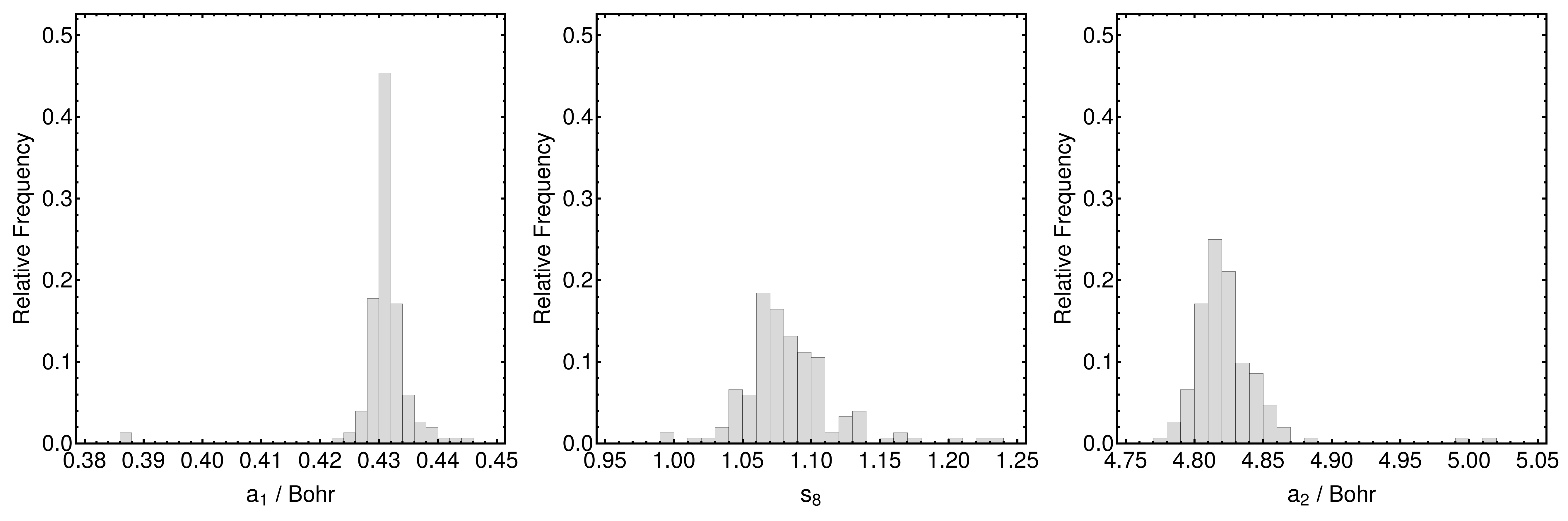}
\end{center}
\caption{\label{fig:jackknife}\small Relative frequency of the PBE-D3-BJ parameters $a_1$, $s_8$, and $a_2$ as obtained from a jackknife 
analysis.}
\end{figure}

The parameter distributions obtained with the jackknife method are shown in Fig.~\ref{fig:jackknife}. 
As already expected on the basis of the bootstrapping analysis, we find distinct outliers in all three histograms of Fig.~\ref{fig:jackknife}.
For instance, there are two outliers in the distributions of $a_1$ and $a_2$, respectively. 
A closer analysis revealed that for both distributions, these outliers correspond to the jackknife samples where the Rn--Xe dimer and the Xe--Xe from the RG6 set have 
been removed.
Hence, these two dimers heavily bias the original calibration procedure (recall that all compounds of the RG6 set are given an increased weight of 20 in the original 
calibration routine, see Section~\ref{sec:weighting}).
Interestingly, the absence of these dimers and the Rn--Rn dimer led to a significant increase of the MAE with respect to the original reference set as inferred from 
our bootstrap analysis presented above.
While Rn--Xe and Xe--Xe seemed to be important for the parameterization in our bootstrapping analysis (though we noted that this interpretation may have been caused by 
the large weight assigned to these systems), they now appear to bias the parameterization.
Removing the Rn--Xe and Xe--Xe dimers from the original reference set and repeating the optimization, we found the following values for the PBE density functional: 
$a_1$ = 0.3969\,Bohr; $s_8$ = 1.1801; $a_2$ = 5.0241\,Bohr 
The MAE with respect to the original reference set amounts to 0.48\,kcal\,mol$^{-1}$, which is only 0.01\,kcal\,mol$^{-1}$ lower compared to our standard calibration 
procedure.
We emphasize that the MAE is not as large as 0.70\,kcal\,mol$^{-1}$ in this case as it is now given with respect to the new reference data set in which these two dimers 
have been removed.
For the validation sets, we found the following MAEs: HEAVY28: 0.37\,kcal\,mol$^{-1}$; AL2X: 2.24\,kcal\,mol$^{-1}$; DARC: 4.00\,kcal\,mol$^{-1}$. 
These values are slightly larger than for the original parameters, except for DARC, where the MAE is reduced by 0.02\,kcal\,mol$^{-1}$, which is, however, not significant. 
Therefore, even though the original reference set is clearly biased, the resulting parameters are not less reliable than parameters resulting from a parameterization 
where this bias was removed, at least as measured by the validation sets chosen in this study.
For different validation sets, especially those representing chemically different or large molecular systems, the situation could change unpredictably.
We will discuss this issue in detail in Section~\ref{sec:size}.

In general, the fact that a given reference set is skewed indicates that  the chemical space is not represented uniformly well. 
Therefore, for domains that are chemically rather dissimilar to the domain covered by the reference set, significant prediction errors are to be expected.

\subsection{Consideration of Reference Uncertainty}
\label{sec:reference}

Irikura \textit{et al.\@} were the first to discuss the inherent uncertainty of quantum chemical data in detail \cite{irikura2004}.
While a quantum chemical result is effectively reproducible, it is biased due to the introduction of approximations concerning, \textit{e.g.}, the Hamiltonian 
or the construction of the wave function including basis-set truncation.
This bias is generally unknown and can only be accessed through statistical modeling, which provides this bias as an uncertainty.
It is important to examine how robust the D3-BJ parameters are with respect to uncertainties in the reference data. 
To this end, we have repeated the bootstrapping procedure ($B =$ 10'000) for the PBE density functional, but during the generation of the bootstrap samples, we 
replaced the reference energy of each data point by an artificially generated value.
This value is a random number, taken from a Gaussian distribution with the mean of the original reference value, and a standard deviation of 5\,\% of this reference 
value.
Clearly, the choice of a 5\,\% standard deviation is rather arbitrary, but it is intended to reflect the expectation that the reference data employed 
are already of high quality.

\begin{table}[H]
\renewcommand{\baselinestretch}{1.0}
\renewcommand{\arraystretch}{1.0}
\caption{\label{tab:uncertainty}\small{
Comparison of the mean of the empirical parameters, $m(v)$, the standard deviations $s(v)$}, min(MAE), max(MAE), and $m$(MAE) obtained from the original bootstrapping 
procedure (middle column, \textit{cf.\@} Table~\ref{tab:bootstrap-errors}) and a modified bootstrapping procedure where a small uncertainty has been incorporated into the 
reference data (right column).}
\begin{center}
\begin{tabular}{l r r} \hline \hline
                                               & Orig.~Ref.~Data & Modif.~Ref.~Data  \\
\hline 
$m(a_1)$ / Bohr                                   & 0.4191                  & 0.4175                    \\
$m(s_8)$                                          & 1.2367                  & 1.2635                    \\
$m(a_2)$ / Bohr                                   & 4.9545                  & 4.9895                    \\
$s(a_1)$ / Bohr                                & 0.0716                  & 0.0748                    \\
$s(s_8)$                                      & 0.3782                  & 0.3807                    \\
$s(a_2)$ / Bohr                                & 0.2886                  & 0.3123                    \\
min(MAE) / kcal\,mol\textsuperscript{$-$1} & 0.49                    & 0.49                      \\
max(MAE) / kcal\,mol\textsuperscript{$-$1}  & 0.80                    & 0.80                      \\
$m$(MAE) / kcal\,mol\textsuperscript{$-$1}     & 0.50                    & 0.50                      \\
\hline
\hline
\end{tabular}
\renewcommand{\baselinestretch}{1.0}
\renewcommand{\arraystretch}{1.0}
\end{center}
\end{table}

The new parameter values as well as their standard deviations are reported in Table~\ref{tab:uncertainty} and compared to the corresponding values obtained from 
the original bootstrapping analysis (see Section~\ref{sec:bootstrap}). 
None of the empirical parameter values changed significantly upon adding a small uncertainty to the reference data. 
Also the resulting standard deviations are almost unaffected.
Furthermore, the resulting prediction errors (also reported in Table~\ref{tab:uncertainty}) are virtually unchanged. 
Our findings suggest that the D3-BJ parameterization is rather robust with respect to small uncertainties in the reference data employed.

\subsection{Effect of Molecular Size on Model Prediction Uncertainty}
\label{sec:size}

In this section, we investigate how deviations of pairwise dispersion interactions induced by different parameterizations propagate to total dispersion energies.
For this purpose, we introduce some reduced notation for the sake of clarity.
First, we define $E$ to be the total D3-BJ dispersion energy for the \textit{original parameterization},
\begin{equation}
E \equiv E_\text{disp}^\text{D3-BJ} = -\sum_{AB} \sum_{n=6,8} s_n \frac{C_n^{AB}}{R_{AB}^n + \big(a_1 R_0^{AB} + a_2 \big)^n} \ .
\end{equation}
The pairwise dispersion energies are abbreviated as
\begin{equation}
E_n^{AB} \equiv -\frac{C_n^{AB}}{R_{AB}^n + \big(a_1 R_0^{AB} + a_2 \big)^n} \ ,
\end{equation}
which sum up to the total dipole--dipole ($n = 6$) and dipole--quadrupole ($n = 8$) interaction terms
\begin{equation}
E_n \equiv \sum_{AB} E_n^{AB} \ .
\end{equation}
Since $s_6 = 1$, the total dispersion energy for the original parameterization reads
\begin{equation}
\label{eq:edisp_symbol}
E = E_6 + s_8E_8 \ .
\end{equation}
Note that $E_6$ and $E_8$ are system dependent and that we do not indicate this dependence to keep the notation uncluttered.

Now, we determine the implications of a modified parameterization on the pairwise and total dispersion energies.
We begin with the empirical parameters of the BJ damping function, $a_1$ and $a_2$.
We express the new parameter $a_{1,b}$ obtained from another calibration (\textit{e.g.}, against a new reference set indicated by the index variable $b$) as the 
sum of the original parameter $a_1$ and a sample-dependent parameter $d_b$,
\begin{equation}
\label{eq:a1b}
a_{1,b} \equiv a_1 + d_b \ .
\end{equation}
Similarly, we define the new parameter $a_{2,b}$ as the sum of $a_2$ and the product of $d_b$ and another sample-dependent parameter $\varepsilon_b$,
\begin{equation}
\label{eq:a2b}
a_{2,b} \equiv a_2 + \varepsilon_b d_b \ .
\end{equation}
As we have seen in the preceding sections, both $a_1$ and $a_2$ are always larger than zero. 
Consequently, we obtain the boundary conditions $d_b > -a_1$ and $\varepsilon_b > -a_2/d_b$ for the sample-dependent parameters.
The empirical damping term $a_1R_0^{AB} + a_2$ changes accordingly for a different parameterization,
\begin{equation}
\label{eq:empirical_term}
a_{1,b}R_0^{AB} + a_{2,b} = \underbrace{a_1R_0^{AB} + a_2}_{\equiv p^{AB}, \ \text{original term}} + \underbrace{d_b \big(R_0^{AB} + \varepsilon_b \big)}_{\equiv q_b^{AB}} \ ,
\end{equation}
and can be written as the sum of the original empirical damping term, $p^{AB}$, and a sample-dependent term, $q_b^{AB}$.
With this definition, we can determine some general properties that hold for all pairwise interactions under certain conditions; $\forall AB$ holds
\begin{equation}
\label{eq:general_properties}
p^{AB}
\begin{cases}
< p^{AB} + q_b^{AB} \ \text{if} \ \varepsilon_b > -\min\big(R_0^{AB}\big) \ , \ d_b > 0 \ , \ \text{or} \ \varepsilon_b < -\max\big(R_0^{AB}\big) \ , \ d_b < 0 \\
> p^{AB} + q_b^{AB} \ \text{if} \ \varepsilon_b > -\min\big(R_0^{AB}\big) \ , \ d_b < 0 \ , \ \text{or} \ \varepsilon_b < -\max\big(R_0^{AB}\big) \ , \ d_b > 0 
\end{cases} \ .
\end{equation}
For the pairwise dispersion energies arising from the parameterization of the $b$-th reference set,
\begin{equation}
E_{n,b}^{AB} \equiv -\frac{C_n^{AB}}{R_{AB}^n + \big(p^{AB} + q_b^{AB} \big)^n} \ ,
\end{equation}
an analogous analysis is possible: $\forall AB$ we have
\begin{equation}
\label{eq:general_properties_alias}
E_n^{AB}
\begin{cases}
< E_{n,b}^{AB} \ \text{if} \ \varepsilon_b > -\min\big(R_0^{AB}\big) \ , \ d_b > 0 \ , \ \text{or} \ \varepsilon_b < -\max\big(R_0^{AB}\big) \ , \ d_b < 0 \\
> E_{n,b}^{AB} \ \text{if} \ \varepsilon_b > -\min\big(R_0^{AB}\big) \ , \ d_b < 0 \ , \ \text{or} \ \varepsilon_b < -\max\big(R_0^{AB}\big) \ , \ d_b > 0 
\end{cases} \ .
\end{equation}
Note here that all pairwise dispersion energies are strictly negative quantities.
This analysis reveals that for a significant domain of $\varepsilon_b$, the deviation between $E_n$ and
\begin{equation}
E_{n,b} \equiv \sum_{AB} E_{n,b}^{AB}
\end{equation}
will increase with \textit{every} pairwise dispersion interaction, \textit{i.e.}, the difference between these two quantities,
\begin{equation}
\label{eq:Delta}
\Delta_{n,b} \equiv E_{n,b} - E_n < -E_n \ ,
\end{equation}
will constantly increase with an increasing number of pairwise interaction terms and, hence, for increasing sizes of molecular systems.

\begin{figure}[t]
\begin{center}
\includegraphics[width=0.6\textwidth]{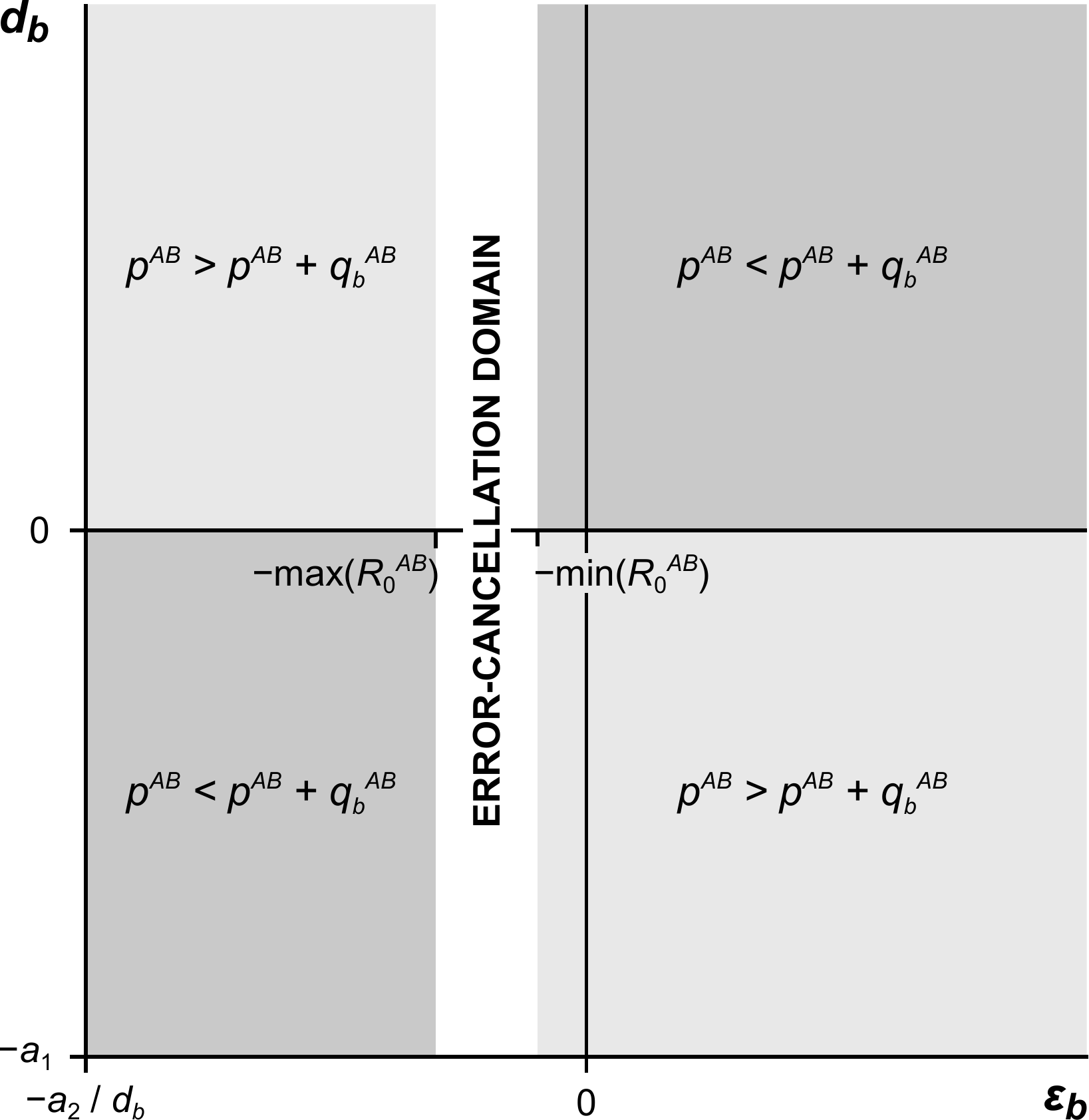}
\end{center}
\caption{\label{fig:error_compensation}\small 
Effect of a new parameterization of the BJ damping function on its original empirical damping term, $p^{AB} = a_1R_0^{AB}+a_2$.
The new empirical damping term is defined as $p^{AB} + d_b\big(R_0^{AB} + \varepsilon_b \big)$.
If $\varepsilon_b > -\min\big(R_0^{AB}\big) \ , \ d_b > 0$ or $\varepsilon_b < -\max\big(R_0^{AB}\big) \ , \ d_b < 0$, the new empirical damping term will be larger 
than the original one for every pairwise interaction (dark-gray domain).
If $\varepsilon_b > -\min\big(R_0^{AB}\big) \ , \ d_b < 0$ or $\varepsilon_b < -\max\big(R_0^{AB}\big) \ , \ d_b > 0$, the new empirical damping term will be smaller 
than the original one for every pairwise interaction (light-gray domain).
Only if $-\max\big(R_0^{AB}\big) < \varepsilon_b < -\min\big(R_0^{AB}\big)$, error cancellation may occur, \textit{i.e.}, some empirical damping terms will increase 
whereas others will decrease.}
\end{figure}

Error cancellation with respect to the original parameterization can only occur in two ways.
First, the condition $-\max\big(R_0^{AB}\big) < \varepsilon_b < -\min\big(R_0^{AB}\big)$ is fulfilled as illustrated in Fig.~\ref{fig:error_compensation} (we will refer to
this a type-I error cancellation below).
For a diatomic system, the white error-cancellation domain in Fig.~\ref{fig:error_compensation} reduces to a single value, $R_0^{AB}$, which depends on the atomic 
numbers of the two atoms $A$ and $B$.
This example illustrates that the error-cancellation domain is system-dependent, \textit{i.e.}, its position and width will be generally different for different molecular systems.
Consequently, if error cancellation may occur for one molecular system, it may not be found for another system.
Moreover, the further the distance between $\varepsilon_b$ and the error-cancellation domain and the larger the absolute value of $d_b$, the larger 
$\vert \Delta_{n,b} \vert$ will become.

The second possibility for error cancellation (type-II error cancellation) depends on the $s_8$ parameter in the $b$-th reference set,
\begin{equation}
s_{8,b} \equiv s_8 + \delta_b \ ,
\end{equation}
where we can postulate $\delta_b > -s_8$ without loss of generality since dispersion energies are strictly negative quantities.
When rearranging the expression for the total dispersion energy corresponding to the $b$-th parameterization,
\begin{eqnarray}
\label{eq:edisp_b}
E_b \equiv E_{6,b} + s_{8,b}E_{8,b} =& E_6 + s_8E_8 &+ \ \Delta_{6,b} + s_8\Delta_{8,b} + \delta_bE_{8,b}  \nonumber \\
=& E &+ \ \Delta_{6,b} + s_8\Delta_{8,b} + \delta_bE_{8,b} \ ,
\end{eqnarray}
we find that $E_b$ is the sum of the original total dispersion energy, $E$, the change in the total dipole--dipole interaction, $\Delta_{6,b}$, the change in the 
total dipole--quadrupole interaction, $s_8\Delta_{8,b}$, and the product of the change in $s_8$, $\delta_b$, and the total dipole--quadrupole interaction of the 
$b$-reference set, $E_{8,b}$.
If $\varepsilon_b$ does not lie in the error-cancellation domain, both $\Delta_{6,b}$ and $s_8\Delta_{8,b}$ will be either strictly positive or negative; and since 
$E_{8,b}$ is always strictly negative, error cancellation can only occur if $\delta_b$ has the same sign as $\Delta_{n,b}$ ($n$ = 6, 8).
On the contrary, if $\varepsilon_b$ does lie in the error-cancellation domain, $\Delta_{6,b}$ and $s_8\Delta_{8,b}$ will have smaller absolute values or even 
different signs, and the importance of $\delta_b$ as an error-cancellation parameter decreases.
Still, only if $\delta_b$ has the same sign as $\Delta_{6,b} + s_8\Delta_{8,b}$, further error cancellation is possible.
At the same time, no matter what value $\varepsilon_b$ takes, $\delta_b$ can also have an error-enhancing effect, \textit{i.e.}, if it shows the opposite sign to 
$\Delta_{6,b} + s_8\Delta_{8,b}$.

We emphasize that all terms in Eq.~(\ref{eq:edisp_b}) are system-dependent and, therefore, the accumulation of error needs to be studied case by case.
Clearly, for a set of small molecular systems, the error spread can be expected to be smaller than for a set of large systems.
We understand that the accumulation of error is a general property of models building on sums of pairwise interactions and is not a result of the specific 
formulation of the D3 approach or any other semiclassical dispersion correction.
Rather, the success of such models depends on the construction of the pairwise interaction terms themselves\,---\,assuming that a sum over pairwise interactions 
is the dominant contribution to the total energy.
The more accurately these terms describe the underlying physical effects, the smaller will be the expected error per term.

To develop an intuition of how strongly errors may accumulate with an increasing number of pairwise D3-BJ interaction terms, we will first study the correlations 
between the three empirical parameters $a_1$, $s_8$, and $a_2$, which are reported in Table~\ref{tab:correlation}.
The correlation between two generic parameters $v$ and $w$ is defined as the covariance of $v$ and $w$ divided by the product of the standard deviations of $v$ and $w$,
\begin{equation}
\label{eq:corr}
r(v,w) = \frac{s^2(v,w)}{s(v)s(w)} \ .
\end{equation}

\begin{table}[H]
\renewcommand{\baselinestretch}{1.0}
\renewcommand{\arraystretch}{1.0}
\caption{\label{tab:correlation}\small 
Correlation matrix of the bootstrapped PBE-D3-BJ parameters $a_1$, $s_8$, and $a_2$.
See Eq.~(\ref{eq:corr}) for a definition of the correlation matrix.
}
\begin{center}
\begin{tabular}{l r r r } \hline \hline
 & $a_1$          & $s_8$          &  $a_2$  \\
\hline 
$a_1$ & 1.00 & 0.51 & $-$0.74 \\
$s_8$ & 0.51 & 1.00 & 0.17 \\
$a_2$ & $-$0.74 & 0.17 & 1.00 \\
\hline
\hline
\end{tabular}
\renewcommand{\baselinestretch}{1.0}
\renewcommand{\arraystretch}{1.0}
\end{center}
\end{table}

It follows from Eq.~(\ref{eq:general_properties}) that the first type of error cancellation can only occur if $\varepsilon_b < 0$ (see also Fig.~\ref{fig:error_compensation}).
In these cases, \textit{cf.\@} Eqs.~(\ref{eq:a1b})~and~(\ref{eq:a2b}),  $a_1$ increases (decreases) and $a_2$ decreases (increases).
Consequently, our bootstrapped negative correlation between $a_1$ and $a_2$, $r(a_1,a_2) = -0.74$, indicates a tendency for type-I error cancellation.
Note however, that an increase (decrease) in $a_1$ coinciding with a decrease (increase) in $a_2$ does not imply error cancellation.

Next, we find that the correlation between $a_1$ and $s_8$ is relatively large and positive, $r(a_1,s_8) = 0.51$, compared to the positive correlation between 
$a_2$ and $s_8$, $r(a_2,s_8) = 0.17$.
Given that $\Delta_{6,b} + s_8\Delta_{8,b}$ is positive, which indicates a tendency for an increase in $a_2$ whereas no direct tendency can be derived for the 
change in $a_1$.
However, based on the negative correlation between $a_1$ and $a_2$ which we estimated through bootstrapping, we expect a decrease in $a_1$ on average if 
$\Delta_{6,b} + s_8\Delta_{8,b} > 0$.
In this case, type-II error cancellation can only occur if $\delta_b$ is also positive.
Since we expect $a_1$ to be decreasing, we also expect $s_8$ to be decreasing due to the positive correlation between $a_1$ and $s_8$, which would however, imply, 
$\delta_b < 0$.
Hence, we do not find direct evidence for a tendency towards type-II error cancellation.

Clearly, error accumulation does not occur uniformly over all pairwise dispersion interactions.
If the distance between two atoms is rather large, we expect the possible error in the corresponding interaction term to be smaller than for two atoms that are closer.
Therefore, one should expect the sampled deviation in the total dispersion energy to be a function of some effective system size, $\Omega_\text{eff}$, which we define here as
\begin{equation}
\Omega_\text{eff} = \sum_{AB} \frac{f^{(6)}_\text{damp,BJ}(R_{AB})}{R_{AB}^6} = \sum_{AB} \frac{1}{R_{AB}^6 + \big(a_1R_0^{AB}+a_2\big)^6} \ ,
\end{equation}
where $a_1$, $s_8$, and $a_2$ are the parameters obtained from calibration against the original reference set \cite{grimme2011}.
With this definition, the effective system size is a monotonically decreasing function of $R_{AB}$ that asymptotically approaches zero. Note that
with this definition, the resulting effective size values are comparatively small.

\begin{figure}[H]
\begin{center}
\includegraphics[width=\textwidth]{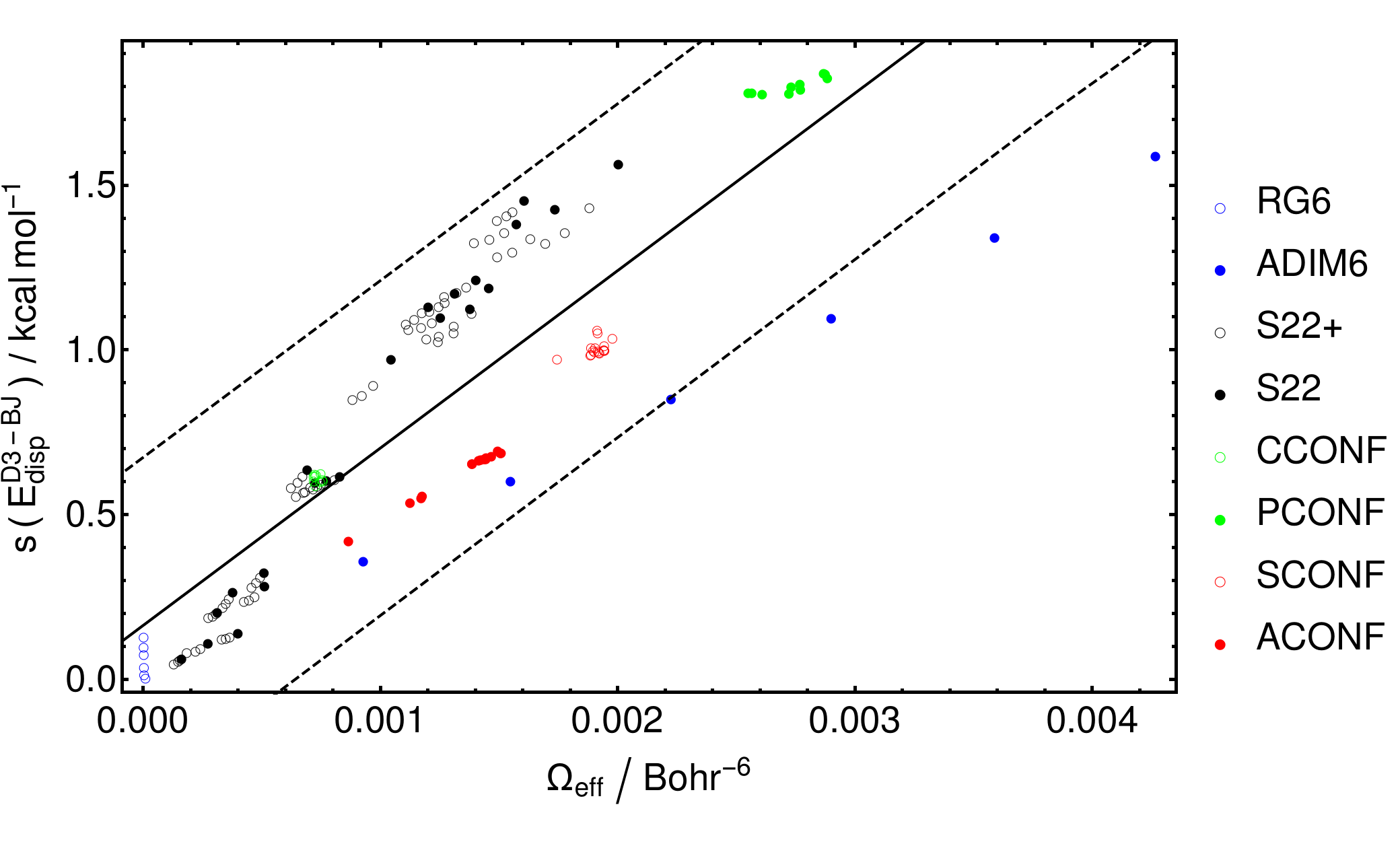}
\end{center}
\caption{\label{fig:effective_size}\small Bootstrapped standard deviation of the total dispersion energy, $s\big(E_\text{disp}^\text{D3-BJ}\big)$, as a function 
of the effective system size, $\Omega_\text{eff}$, for the reference subsets \cite{grimme2011} employed in the parameterization of the D3-BJ model.
The solid line represents the least-squares fit to the data points, whereas the dashed lines represent the two-sided 95\,\% confidence interval.
}
\end{figure}

In Fig.~\ref{fig:effective_size}, the bootstrapped standard deviation of the total dispersion energy, $s\big(E_\text{disp}^\text{D3-BJ}\big)$, is shown as a 
function of the effective system size, $\Omega_\text{eff}$.
This and all following results were produced with the PBE-D3-BJ model.
We find a significant increasing trend, which is clearly linear for most of the individual reference subsets, \textit{e.g.}, for the homologous series of alkanes 
(ADIM6), we find maximum correlation between $s\big(E_\text{disp}^\text{D3-BJ}\big)$ and $\Omega_\text{eff}$ (perfect linearity).
The increase of $s\big(E_\text{disp}^\text{D3-BJ}\big)$ per unit $\Omega_\text{eff}$ is 366.59\,kcal\,mol$^{-1}$ for the ADIM6 set and is the smallest among all reference 
subsets.
For the S22 and S22+ sets, which comprise many unsaturated organic molecules, the increase of $s\big(E_\text{disp}^\text{D3-BJ}\big)$ per unit $\Omega_\text{eff}$ amounts 
already to 927.50\,kcal\,mol$^{-1}$.
Averaged over all reference subsets, we find a slope of $\Delta s\big(E_\text{disp}^\text{D3-BJ}\big) / \Delta \Omega_\text{eff} = 538.76$\,Bohr$^{6}$\,kcal\,mol$^{-1}$
with a negligible intercept of $0.16$\,kcal\,mol$^{-1}$.
Hence, for a molecular system with an effective size of 0.01\,Bohr$^{-6}$~(see below for an example molecule of approximately this size), the effect of parameter 
uncertainty on the MPU may already be on the order of 5\,kcal\,mol$^{-1}$.

\begin{figure}[H]
\begin{center}
\includegraphics[width=0.6\textwidth]{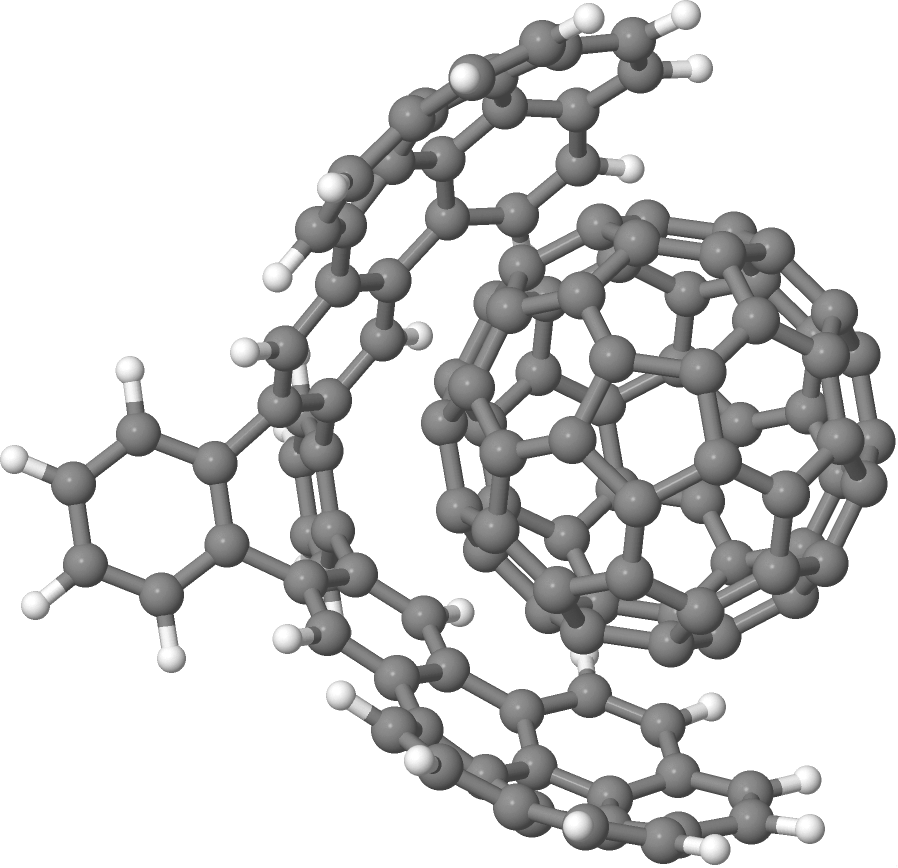}
\end{center}
\caption{\label{fig:bucky}\small Structure model of the C$_{60}$ buckycatcher (molecular coordinates were taken from Ref.~\cite{phys_chem_chem_phys_2008_10_2813}).}
\end{figure}

We will now extrapolate these findings to a specific example with an effective size that surmounts all 152 systems contained in the original 
reference set.
For this purpose, we selected the C$_{60}$ ``buckycatcher'', that has been a prototypical example for the development of dispersion corrections\cite{grimme2010},
with an effective size of $\Omega_\text{eff} = 1.221$\,$\times 10^{-2}$\,Bohr$^{-6}$\ (Fig.~\ref{fig:bucky}) and a total dispersion energy of $E_\text{disp}^\text{D3-BJ} = -229.27$\,kcal\,mol$^{-1}$ 
as determined with our parameterization provided in Table~ \ref{tab:param-bj}.
Based on the slope and intercept obtained for the entire reference set, we would expect a standard deviation of the total dispersion energy of 
$s\big(E_\text{disp}^\text{D3-BJ}\big) = 6.74$\,kcal\,mol$^{-1}$.
However, when calculating $E_\text{disp}^\text{D3-BJ}$ for all 10'000 bootstrapped parameter sets, we found a much larger standard deviation of 
$s\big(E_\text{disp}^\text{D3-BJ}\big) = 14.08$\,kcal\,mol$^{-1}$.
As the buckycatcher is a highly unsaturated system, we expect its uncertainty in the total dispersion energy to rather follow the trend exhibited by 
the S22 and S22+ subsets.
With this assumption, we obtained $s\big(E_\text{disp}^\text{D3-BJ}\big) = 11.23$\,kcal\,mol$^{-1}$, which is more in line with the result of our bootstrap analysis.
Noteworthy, the maximum bootstrap spread in $E_\text{disp}^\text{D3-BJ}$, $\max(E_{n,b}) - \min(E_{n,b})$, even amounts to 176.90\,kcal\,mol$^{-1}$.

In Section~\ref{sec:weighting}, we found that the original and uniform weighting schemes yield D3-BJ parameters for PBE that are significantly different from each other.
In the case of the buckycatcher, the deviations in these parameter sets amount to a difference in $E_\text{disp}^\text{D3-BJ}$ of 
63.47\,kcal\,mol$^{-1}$, which is certainly too large to provide a quantitative description of dispersion interactions in that molecular system based on the D3-BJ model.
Contrary to the previous scenario, we find a small difference of 4.95\,kcal\,mol$^{-1}$ for the skewed and balanced reference data sets (\textit{cf.\@} 
Section~\ref{sec:jackknife}).
These findings indicate that the weighting scheme has a more significant effect on the MPU than small modifications of the reference set.
Furthermore, when comparing $E_\text{disp}^\text{D3-BJ}$ for the mean parameter sets obtained from the original and modified bootstrapping procedures 
(\textit{cf.\@} Section~\ref{sec:reference}), we find a difference of only 1.91\,kcal\,mol$^{-1}$.

Another factor determining the calibration procedure is the representation of residuals in the cost function.
Most commonly, the absolute value of the residual and the squared residual are employed (\textit{cf.\@} Section~\ref{sec:params}). 
While the conclusions drawn in this paper are not affected by the choice of the specific residual representation, it certainly has a numerical impact on 
the dispersion energy since different representations (and, hence, cost functions) generally yield different optimal parameter values.

\section{Conclusions}
\label{sec:conclusion}

In this work, we have subjected D3-type semiclassical dispersion corrections to a rigorous statistical analysis, with a focus on D3-BJ.
This study is the first assessing D$x$-type semiclassical dispersion interactions based on a rigorous statistical analysis.
We have recalibrated the empirical parameters $s_{r,6}$ and $s_8$ for D3-zero as well as $a_1$, $s_8$, and $a_2$ for D3-BJ.
Overall, our results agree well with the original parameters reported by Grimme and coworkers\cite{grimme2010}; in some cases, our parameterization results in a 
slightly lower mean prediction error compared to the original one, and in other cases, we arrive at the opposite result.
We therefore do not intend to suggest our new D3-BJ parameterization as an improvement over the original D3-BJ parameterization.

For the molecular systems contained in the reference set, these slightly different parameterizations only have a small effect on the MAE of Eq.~(\ref{eq:mad}) 
(less than 1\,kcal\,mol$^{-1}$). However, we studied in detail the effect of different parameterizations on increasingly larger molecular systems 
motivated by the fact that physical models based on sums of constantly attractive pairwise interactions are particularly prone to error accumulation.
We examined this issue for a fairly large range of molecular sizes from two angles.
First, we studied the effect of the specific calibration procedure (choice of reference set and cost function, and consideration of uncertainty in the reference data) 
on the dispersion energy.
Second, we also took into account the effect of uncertainty in the empirical D3 parameters on the dispersion energy.
For both, we carried out a rigorous analysis of error accumulation for the three empirical D3-BJ parameters.
While the error introduced by two different parameterizations can partially cancel out, we found that this effect is system-dependent, suggesting that error cancellation 
may occur in some molecules, but be completely absent in others (\textit{i.e.}, every pairwise interaction resulting from one parameterization is constantly larger (or 
smaller) compared to another parameterization).
In that case, the error in the dispersion energy may become dramatically large as it would monotonically increase with the number of pairwise interactions and, hence, 
with molecular size.
Although the correlations of the empirical D3-BJ  parameters obtained from our statistical analysis suggest that partial error compensation is likely to occur,
we found that error cancellation of individual pairwise interaction terms is not possible over wide parameter ranges.

We find that the nonuniform weighting scheme originally employed by Grimme and coworkers\cite{grimme2010} sometimes leads to larger errors (as assessed through a variety of different validation sets) than those that would result from a uniform weighting scheme that considers all data points on the same footing. 
When considering a relatively large molecular system\,---\,here, the prominent C$_{60}$ buckycatcher\,---\,different weighting schemes lead to deviations in the dispersion energy of more than 60\,kcal\,mol$^{-1}$, which is clearly too large for a quantitative description of the total dispersion energy.
Furthermore, we examined the effect of biased reference data on the dispersion energy.
When assuming a small uncertainty in the reference data of 5\,\%, we find only slightly different parameters, the deviation of which accumulates to
about 5\,kcal\,mol$^{-1}$ for the buckycatcher and less than or about 1\,kcal\,mol$^{-1}$ for all other molecular systems studied.

Next, we investigated the role of parameter uncertainty for error accumulation.
Here, parameter uncertainty was estimated on the basis of a nonparametric bootstrap analysis.
It turns out that the standard deviations of the means of the empirical parameters are small in all cases.
Moreover, the bootstrap analysis confirms the MAE to be a reliable estimate of prediction uncertainty.
Most interestingly, based on the bootstrap analysis we found a linear dependence of the expected energy uncertainty on the effective molecular size.
For all molecules contained in the reference set, this uncertainty is smaller than 2\,kcal\,mol$^{-1}$, while for the buckycatcher, the small estimated parameter 
uncertainty leads to a standard deviation in the dispersion energy of 14\,kcal\,mol$^{-1}$ (however, the maximum deviation in the dispersion energy determined with bootstrapping amounts to a surprisingly large value of almost 180\,kcal\,mol$^{-1}$
in this case).

We also applied the jackknife method, which is related to the bootstrap approach, to examine the consistency of the reference data.
Our analysis shows that the original reference set employed by Grimme \textit{et al.\@} is slightly biased, even though our results also suggest that the effect of this bias on the dispersion energy is negligible in the cases studied.

To support routine DFT-D3 calculations in efficiently estimating uncertainties of dispersion energies, we will make the program \textsc{BootD3} available on our web page.
\textsc{BootD3} reads a user-specified file of Cartesian coordinates and runs Grimme's D3 implementation\cite{dftd3_website} for each of the 10'000 bootstrapped parameter sets generated for this study.
In this way, a virtual error bar can be assigned to a dispersion energy.
All density functionals examined in this work and the corresponding bootstrapped parameter sets are available in \textsc{BootD3}.

\section*{Acknowledgments}

This work was financially supported by the Schweizerischer Nationalfonds (Project No.~200020\_169120).

\providecommand{\refin}[1]{\\ \textbf{Referenced in:} #1}




\end{document}